\documentclass[conference,10pt]{IEEEtran}
\IEEEoverridecommandlockouts   
\usepackage[utf8]{inputenc}
\usepackage{newtxtext,newtxmath}
\usepackage[dvipsnames]{xcolor}
\usepackage[T1]{fontenc}
\usepackage{array}
\usepackage{url}
\usepackage{balance}
\usepackage{hyperref}
\usepackage{microtype}
\usepackage{graphicx}
\usepackage[binary-units,per-mode=symbol]{siunitx}
\usepackage{float}
\usepackage{lipsum}
\usepackage{enumitem}

\ifCLASSOPTIONcompsoc
  \usepackage[caption=false, font=normalsize, labelfont=sf, textfont=sf]{subfig}
  \usepackage[nocompress]{cite}
\else
  \usepackage[caption=false, font=footnotesize]{subfig}
  \usepackage{cite}
\fi

\usepackage[acronym,nohypertypes=acronym]{glossaries}
\usepackage[capitalise]{cleveref}

\setacronymstyle{long-short}

\IEEEtriggercmd{\balance}
\IEEEtriggeratref{20}

\newcommand{\papertitle}{Social Distancing and the Internet: What Can Network Performance Measurements Tell Us?}

\hypersetup{
  colorlinks=true,
  linkcolor=black,
  urlcolor=black,
  citecolor=black,
  pdftitle={\papertitle}
}

\newacronym{FCC}{FCC}{Federal Communications Commission}
\newacronym{MBA}{MBA}{Measuring Broadband America}
\newacronym{ISP}{ISP}{internet service provider}
\newacronym{IXP}{IXP}{internet exchange point}
\newacronym{MNO}{MNO}{mobile network operator}
\newacronym{CSV}{CSV}{comma-separated values}
\newacronym{UDP}{UDP}{User Datagram Protocol}
\newacronym{ICMP}{ICMP}{Internet Control Message Protocol}
\newacronym{VoIP}{VoIP}{voice over IP}
\newacronym{DNS}{DNS}{Domain Name System}
\newacronym{SQL}{SQL}{Structured Query Language}
\newacronym{VM}{VM}{virtual machine}
\newacronym{LAN}{LAN}{local area network}
\newacronym{FIPS}{FIPS}{Federal Information Processing Standard}
\newacronym{UTC}{UTC}{Coordinated Universal Time}
\newacronym{CDF}{CDF}{cumulative distribution function}
\newacronym{VPN}{VPN}{virtual private network}
\newacronym{DIMACS}{DIMACS}{Center for Discrete Mathematics \& Theoretical Computer Science}
\newacronym{LA}{LA}{Los Angeles}
\newacronym{US}{US}{United States}
\newacronym{UK}{UK}{United Kingdom}
\newacronym{IP}{IP}{Internet Protocol}
\newacronym{NCTA}{NCTA}{Internet \& Television Association}
\newacronym{DSL}{DSL}{digital subscriber line}
\newacronym{EPA}{EPA}{U.S. Environmental Protection Agency}
\newacronym{MSA}{MSA}{metropolitan statistical area}

\begin{document}
\title{\papertitle}

\author{
    \IEEEauthorblockN{
        \href{mailto:jdm2240@barnard.edu}{\color{black}Jessica De Oliveira Moreira}\IEEEauthorrefmark{1},
        \href{mailto:amey.pasarkar@columbia.edu}{\color{black}Amey Praveen Pasarkar}\IEEEauthorrefmark{2},
        \href{mailto:wc2741@columbia.edu}{\color{black}Wenjun Chen}\IEEEauthorrefmark{2},
        \href{mailto:wh2453@columbia.edu}{\color{black}Wenkai Hu}\IEEEauthorrefmark{2},\\
        \href{mailto:janakj@cs.columbia.edu}{\color{black}Jan Janak}\IEEEauthorrefmark{3},
        \href{mailto:hgs@cs.columbia.edu}{\color{black}Henning Schulzrinne}\IEEEauthorrefmark{3}
    }
    \IEEEauthorblockA{%
        \IEEEauthorrefmark{1}Barnard College, New York, USA
    }
    \IEEEauthorblockA{%
        \IEEEauthorrefmark{2}Fu Foundation School of Engineering and Applied Science, Columbia University, New York, USA
    }
    \IEEEauthorblockA{%
        \IEEEauthorrefmark{3}Department of Computer Science, Columbia University, New York, USA
    }%
    Email:
    \href{mailto:jdm2240@barnard.edu}{\color{black}jdm2240@columbia.edu},
    \{%
      \href{mailto:amey.pasarkar@columbia.edu}{\color{black}amey.pasarkar},%
      \href{mailto:wc2741@columbia.edu}{\color{black}wc2741},%
      \href{mailto:wh2453@columbia.edu}{\color{black}wh2453}%
    \}@columbia.edu,
    \{%
      \href{mailto:janakj@cs.columbia.edu}{\color{black}janakj},%
      \href{mailto:hgs@cs.columbia.edu}{\color{black}hgs}%
    \}@cs.columbia.edu%
    \thanks{Jessica De Oliveira Moreira, Amey Praveen Pasarkar, Wenjun Chen, and Wenkai Hu contributed equally to this research. The work of Jessica De Oliveira Moreira was supported by Craig Newmark Philanthropies. Corresponding authors: Jan Janak; Henning Schulzrinne.}
}

\maketitle

\begin{abstract}
The COVID-19 pandemic and related restrictions forced many to work, learn, and socialize from home over the internet. There appears to be consensus that internet infrastructure in the developed world handled the resulting traffic surge well. In this paper, we study network measurement data collected by the \glsdesc{FCC}['s] \glsdesc{MBA} program before and during the pandemic in the \gls{US}. We analyze the data to understand the impact of lockdown orders on the performance of fixed broadband internet infrastructure across the US, and also attempt to correlate internet usage patterns with the changing behavior of users during lockdown. We found the key metrics such as change in data usage to be generally consistent with the literature. Through additional analysis, we found differences between metro and rural areas, changes in weekday, weekend, and hourly internet usage patterns, and indications of network congestion for some users.


\end{abstract}


\glsresetall

\section{Introduction}\label{sec:introduction}

Much has been written about whether the residential internet infrastructure in the \gls{US} can cope with the sudden conversion of work into tele-work, school into distance learning and movies into Netflix binging, albeit usually based on data of dubious quality. This topic has seen renewed interest in 2020 when the COVID-19 pandemic suddenly forced a large percentage of the \gls{US} population to work, learn, and socialize from home over the internet for extended periods of time.

Network measurements based on data from a variety of vantage points tell us about the performance of internet infrastructure. However, such measurements may also tell us about the effort and success of social distancing, mobility trends and patterns, the ability of different communities to move to internet-mediated modes of interaction and their longer-term sustainability.

We attempt to address both questions in this paper. Drawing on the \gls{FCC}['s] \gls{MBA} program dataset~\cite{mba}, available longitudinally since 2011, we estimate the impact of increased internet usage during the pandemic on a variety of performance metrics. We specifically look at throughput and packet loss, disaggregating by coarse-grained geography, population density, and time period. Even though it usually does not draw much attention, we are particularly interested in upstream performance, as video conferencing has increased greatly during the pandemic.

At least as important as measuring the impact of COVID-19 on internet performance is using indirect metrics, such as data usage, to estimate the success of social distancing measures, while maintaining user privacy. We draw on two metrics to explore whether we can detect the daytime and evening usage of residential internet users, indicating work from home or distance learning, as well as observance of stay-at-home orders on weekends.

The \gls{MBA} program records user network data usage and suspends active measurements during periods of significant network activity, such as video conferencing. Thus, we might be able to deduce when users are actively using their fixed broadband connection and whether such usage reflects, e.g., that restrictions were imposed earlier in Washington State and the New York metro area.

Our key findings can be summarized as follows.
\begin{itemize}[leftmargin=0.15in]
    \item We find a significant increase in data usage, consistent with lockdown timeline and observations made by other researchers.
    \item The traffic spike in rural areas is initially shorter and less pronounced than in metro areas, however, we see the trend start to reverse by June 2020.
    \item Our data suggest that users in metro areas react to government recommendations faster than users in less populated areas.
    \item We see an increase in daytime weekend traffic from March to May 2020. By June the levels return to their pre-lockdown levels, suggesting that users might have either complied with stay-at-home orders only initially, or that they spent more time (socially-distanced) outside.
    \item Consistent with others, we see the weekday daytime traffic pattern approximating the weekend pattern.
    \item We see a 5\% decrease in average speed for about 60-70\% of test units and also found a significant (almost 50\%) increase in packet loss for some test units in mid-March 2020, suggesting that increased internet usage may have caused network congestion.
\end{itemize}

Our study is based solely on the \gls{FCC} \gls{MBA} dataset which has known limitations. For example, it only covers the largest \glspl{ISP} and omits fixed wireless and satellite providers, the latter by their choice. We also briefly discuss how more systematic data gathering during both short-term natural disasters and long-term epidemics could supplement our understanding of how households deal with such crises, providing information to policy makers in near real-time and in a privacy-respecting manner.

Several authors have published similar studies. The studies are based on data from network services~\cite{bottger2020internet,akamai,facebook}, \glspl{ISP} and \glspl{IXP}~\cite{feldmann2020lockdown,liu2020characterizing}, or mobile networks~\cite{lutu2020characterization}. Our work aims to complement such efforts. We study similar metrics using data collected by \gls{US}[-based] fixed broadband internet users participating in the \gls{FCC} \gls{MBA} program.

The rest of the paper is organized as follows. We review related work on COVID-19 impacts on the internet in \cref{sec:related-work}. The datasets used in all our analyses are discussed in \cref{sec:datasets}. We describe our methodology in \cref{sec:methodology}. In \cref{sec:data-usage-by-population-demographics} we examine broadband internet usage during lockdown in metropolitan and rural counties. \cref{sec:hourly-data-usage-patterns} studies hourly data usage patterns before and during lockdown. In \cref{sec:average-speed}, we look at the effects of COVID-19 lockdown on fixed broadband upload and download speeds. \cref{sec:packet-loss-analysis} examines packet loss during lockdown. We discuss our findings in \cref{sec:discussion} and conclude and discuss future work in \cref{sec:conclusion}.

\section{Related Work}\label{sec:related-work}



A small number of studies focusing on the effects of lockdowns on internet infrastructure have already been published as of November 2020. These studies typically look at traffic trends and patterns from various vantage points. Some look for signs of stress in internet infrastructure, while others attempt to correlate traffic data with user behavior during lockdowns.

Feldmann et al.~\cite{feldmann2020lockdown} studied the effects of lockdowns on internet traffic using traffic data obtained from several \glspl{ISP} and \glspl{IXP} in Europe and on the \gls{US} East Coast. The authors found that residential traffic increased by 15--20\% after the lockdown and that remote work and education applications saw a 200\% increase in traffic. The regular weekday traffic pattern transformed into a weekend-like pattern with a significant traffic increase starting in the morning, from mid-March until roughly mid-May.

B\"{o}ttger et al.~\cite{bottger2020internet} studied similar effects using data from Facebook's global edge network. They observed a traffic increase in the second half of March and found regional correlations between traffic growth and the spread of COVID-19 in the region. The data from Facebook's video services showed a degradation in the quality of user experience in India and South Africa. Authors attributed the degradation to limited capacity and last-mile network congestion in those countries.

Liu et al.~\cite{liu2020characterizing} found a 30--60\% increase in peak traffic volumes in the \gls{US}, a statistically significant increase of 10\% in average latency over two months, and \glspl{ISP} adding interconnect capacity at more than twice the usual rate, based on interconnection data published by participating \glspl{ISP}~\cite{feamster2016revealing}.

Several broadband market reports analyzed the state of broadband infrastructure during the pandemic. OpenVault~\cite{openvault} saw a 40\% broadband traffic increase in Q3 2020 over 2019, but less than 1\% over Q2 2020. Traffic volumes stopped growing in Q3 2020, indicating a new normal for broadband bandwidth usage. Temporarily relaxed data usage quota resulted in faster growth of usage-based accounts compared to flat rate accounts.

Ofcom~\cite{uk-home-broadband-performance} reported that \gls{UK}[-based] broadband service providers were generally able to keep up with the demand in March, with only a 1--2\% decrease in download and upload speeds and 2\% increase in latency.

SamKnows published a series of articles~\cite{samknows-cdn,samknows-video-streaming,samknows-video-conferencing,samknows-usa} about the state of critical infrastructure. Their observations in the \gls{UK} and the \gls{US} showed only slightly worse average download speeds and round trip times during lockdown.

Several network service providers discussed lockdown-related trends and observations on their blogs: Google~\cite{google}, Akamai~\cite{akamai}, Facebook~\cite{facebook}, and Comcast~\cite{comcast}. In all cases, initial traffic surges aligned with mandatory quarantine protocols have been reported, as well as long-term shifts in daily and weekly traffic patterns. The general sentiment appears to be that the existing infrastructure was either able to keep up with the traffic due to over-provisioning, or could be rapidly adapted to meet the increased demand. It is worth mentioning that the above-mentioned network service providers generally report higher traffic surges than broadband monitoring companies like SamKnows or OpenVault.

The \gls{NCTA}~\cite{ncta} reported a usage growth of 30\% (downstream) and 50\% (upstream) since March among its member broadband cable providers. Sandvine~\cite{sandvine} saw a traffic growth of 40\% between February and April. The growth was driven by video, gaming, and social sharing applications which comprise 80\% of all traffic.

Kovacs~\cite{kovacs} compared the performance of fixed broadband internet during the pandemic in the \gls{US} and in Europe. Using data collected by the company Ookla, the author found that \gls{US} fixed broadband networks generally performed better than their comparable European peers, offering 30--35\% higher mean download speed. Kovacs attributed the difference to bigger long term investments in telecommunications infrastructure, more favorable mix of technologies, and lighter-touch regulatory approach in the \gls{US}.

We partially attribute the difference to technology choices. Europe has more \gls{DSL} and less cable, but with strong country-to-country differences. Thus, one would have to compare whether \gls{US} \gls{DSL} also performed worse than \gls{US} cable, for example.



Lutu et al.~\cite{lutu2020characterization} studied traffic patterns obtained from a \gls{UK}[-based] \gls{MNO} and analyzed the effect of mandatory stay-at-home orders on users' mobility behavior. They found an overall decrease in users' mobility following the government’s mandatory stay-at-home orders. The authors found a decrease in cellular data traffic volumes and an overall increase in voice traffic.

\begin{table*}
  \centering
  \caption{Lockdown dates of all 50 \gls{US} states in an increasing date order (left-right, top-bottom) (Source: \url{https://ballotpedia.org})}
  \label{tab:state-lockdown}
  \begin{tabular}{ |c|c|c|c|c|c|c|c|c|c|c|c|c|c|c|c|c|c|c| }
    \hline
    CA   & NY   & IL   & NJ   & CT   & LA   & OH   & OR   & DE   & IN   & MA   & MI   & NM   & VT   & WA   & WV   & HI   & ID   & WI   \\
    \hline
    3/19 & 3/20 & 3/21 & 3/21 & 3/23 & 3/23 & 3/23 & 3/23 & 3/24 & 3/24 & 3/24 & 3/24 & 3/24 & 3/24 & 3/24 & 3/24 & 3/25 & 3/25 & 3/25 \\
    \hline\hline
    CO   & KY   & MN   & NH   & AK   & MT   & RI   & KS   & MD   & NC   & VA   & AZ   & TN   & OK   & NV   & PA   & FL   & ME   & TX   \\
    \hline
    3/26 & 3/26 & 3/27 & 3/27 & 3/28 & 3/28 & 3/28 & 3/30 & 3/30 & 3/30 & 3/30 & 3/31 & 3/13 & 4/1  & 4/1  & 4/1  & 4/2  & 4/2  & 4/2  \\
    \hline\hline
    GA   & MS   & AL   & MO   & SC   & AK   & IA   & NE   & ND   & SD   & UT   & WY   & \multicolumn{7}{|c|}{}                         \\
    \hline
    4/3  & 4/3  & 4/4  & 4/6  & 4/7  & ---  & ---  & ---  & ---  & ---  & ---  & ---  & \multicolumn{7}{|c|}{}                         \\
    \hline
  \end{tabular}
\end{table*}

\section{Datasets}\label{sec:datasets}

The \gls{FCC} runs a program called \glsreset{MBA}\gls{MBA} which collects data about broadband internet usage and performance in the \gls{US}~\cite{mba}. The data are collected by a panel of volunteers with a hardware test unit installed on their fixed broadband internet connection. The test unit is known as the Whitebox and is manufactured by SamKnows~\cite{sam}. The panel consists of 4,000 to 5,000 volunteers (depending on the year), selected by the \gls{FCC} to include 16 major \glspl{ISP}, geographic regions, and broadband connection types~\cite{fcc-report-appendix}. The test unit uses a combination of passive and active measurement techniques to determine the state, parameters, and usage of the internet connection repeatedly throughout the day.

Since 2011, the \gls{FCC} has been publishing annual reports~\cite{mba-studies} on the overall performance of the \gls{US} fixed broadband infrastructure. The reports are based on the data collected from a representative subset of the test units, cross-validated with information provided by fixed broadband \glspl{ISP}. The reports cover approximately 80\% of the \gls{US} population.

In addition, the \gls{FCC} regularly publishes anonymized raw datasets~\cite{data} on its website. All deployed test units generate approximately \SI{20}{\giga\byte} of data per month. In this study, we use the raw datasets to analyze the performance and changing usage patterns of \gls{US} fixed broadband before and during COVID-19 related government-imposed restrictions (lockdowns). We work with the raw data collected from January 2019 to June 2020. At the time of writing, the latest dataset published by the \gls{FCC} was from June 2020.

The raw datasets are published in the form of \gls{CSV} files broken down by the month and type of measurement. The following measurements are included: upload and download speed, the time to fetch a few selected well-known web pages, \gls{UDP} and \gls{ICMP} latency and packet loss when idle and under load, video streaming performance, \gls{VoIP} performance, \gls{DNS} performance, and total bytes uploaded and downloaded by the user. In this project we only use the following raw data subset: upload and download speed, total user bytes, and \gls{UDP}/\gls{ICMP} packet loss.

In addition to raw datasets, we also use the unit profile, unit census, and excluded units datasets as published by the \gls{FCC}. The unit profile dataset provides additional information about the test units, including \gls{ISP} name and service tier, technology, and coarse location. Since some of the information in the unit profile dataset must be validated manually, not all test units are included. The unit census dataset identifies the census block for each test unit. The excluded units dataset lists the test units that were excluded from the most recent annual report. At the time of this study, the most recent unit profile, unit census, and excluded unit datasets were from 2018, i.e., considerably older than the raw datasets. As a result, we must limit some of our analysis to the subset of the test units (and hence users) that have profile and census data.

Analysis that involves population demographic information integrates 2019 U.S. Census Bureau data~\cite{census}. The census data provides county-level population information for all counties in the \gls{US} and can be linked to the \gls{MBA} data through the aforementioned unit profile datasets.

The raw datasets include measurements from 7,539 test units from across the \gls{US}. The number includes all test units that reported at least once between January 2019 and June 2020. The unit profile dataset provides additional information about 4,378 test units. The unit census dataset contains 4,366 test units. The excluded unit dataset contains 1,656 test units. The total size of all datasets used in our analysis was over \SI{300}{\giga\byte}.

By analyzing these datasets, we are aiming to understand not only how the fixed broadband internet infrastructure performed, but also whether we can learn anything about people's changing behavior over the course of the pandemic.

\subsection{Ethical and Privacy Considerations}\label{sec:ethical-and-privacy-considerations}

The \gls{FCC} \gls{MBA} panel consists of volunteers who were informed about the goals and methodology of the program, explicitly opted in, and provided written consent to participate. All publicly released \gls{FCC} \gls{MBA} datasets have any personally identifiable information removed. For example, the unit census dataset which provides a coarse location for each test unit aggregates to blocks with no fewer than 1,000 people for privacy purposes.

All additional datasets used in this study, e.g., the U.S. Census Bureau data, is public information and can be obtained freely from the internet.

\section{Methodology}\label{sec:methodology}

In order to analyze the impacts of COVID-19 lockdowns on fixed broadband internet usage, we need to compare data from time periods before and during lockdown. In the \gls{US}, individual states issued statewide lockdown orders on different dates. \cref{tab:state-lockdown} shows the first date of the statewide lockdown order being enforced for each of the 50 \gls{US} states. The first state to issue a lockdown order was California on March 19. The last statewide lockdown order was issued by South Carolina on April 7. The seven states at the end of the table did not impose lockdown orders.

For the purpose of the analysis presented in this paper, we use January and February 2020 as the pre-lockdown period and mid-March to mid-May as the lockdown period. We consider the period from mid-May until end of June as post-lockdown, i.e., a period where many of the restrictions were being relaxed. To eliminate seasonal effects, we also use data from 2019 to serve as a baseline for comparison in some analysis. Thus, most of our algorithms run on the data from January 2019 to June 2020 (the last dataset published by the \gls{FCC} as of September 2020).

We uploaded the \gls{FCC} \gls{MBA} raw data from January 2019 to June 2020, the 2019 census data, and all the auxiliary datasets (unit profile, unit census, excluded units) into a Google Cloud BigQuery database~\cite{bigquery}. BigQuery allowed us to query the data using a familiar \gls{SQL}[-based] interface.

We then analyzed the data using Google Cloud Datalab~\cite{datalab} running on \glspl{VM} in Google Cloud. Google Cloud Datalab is a version of JupyterLab~\cite{jupyter} with Google Cloud specific extensions (e.g., libraries to access the BigQuery database). All our programs to analyze the data were written in Python 3. We kept everything synchronized and organized in a shared Git repository.

To calculate the total downloaded data per test unit, we sum the number of bytes received from the internet on all wired \glspl{LAN} and Wi-Fi networks. Similarly, to calculate the total uploaded data per test unit, we sum the number of bytes transmitted to the internet from all wired \glspl{LAN} and Wi-Fi networks. In both cases, any data received or transmitted by the test unit itself during active measurements is excluded.

\subsection{Data Cleaning}

The raw dataset includes measurements from all test units that reported at least once within the analyzed time period. We found several test units reporting invalid data, e.g., unrealistic amounts of user traffic. We assumed those test units were malfunctioning and manually removed all their data from the database.

In some of the experiments where a larger number of test units was not necessary, we filtered the raw datasets and included only test units that were not found in the excluded units datasets and that were found in the unit profile dataset. This step gives a smaller test unit subset that roughly includes the test units used in the most recent \gls{FCC} annual report.

\subsection{Population Demographics Preprocessing}

For population demographics analysis, we need to classify test units as belonging to either metropolitan or rural areas. We describe our approach in this section.

The \gls{FCC} \gls{MBA} unit census dataset provides \gls{FIPS} census object identifiers for a subset of the test units in the study. The census object can be one of: block group, tract, county, state. For privacy reasons, the dataset only provides census objects with more than 1,000 people. For example, if a test unit belongs to a census block group with fewer than 1,000 people, a census tract will be used for the test unit instead. If the tract still has fewer than 1,000 people, granularity decreases to county and eventually state level.

Census block groups are designed to contain a specific number of people and therefore provide little information regarding how densely populated an area is. For example, a few blocks in New York City would correspond to a single census block, while an entire town in rural America would also qualify as a census block. Therefore, in order to support an analysis of how users in more densely populated areas behaved relative to those in less densely populated areas, we estimate population density from county-level populations.

\gls{FIPS} census identifiers have a hierarchical structure where the first two digits refer to the state, while the subsequent three refer to the county within the state. Thus, to map test units to populations, we extract the county from each test unit's \gls{FIPS} census identifier and map it to a population using the publicly-available 2019 \gls{US} Census Bureau dataset \cite{census}. In this step, we must exclude approximately 180 test units that have a state-level census identifier only (counties with fewer than 1,000 people).

Among the test units for which we were able to obtain county population, we focus specifically on those that have participated in the study since January 2019 (1,692 test units in total), i.e., we exclude test units that have joined the \gls{FCC} \gls{MBA} program later. This helps ensure that when we compare 2019 and 2020 data, any changes that we observe will be due to external factors like the pandemic and not due to changing test unit sample size.

We classify test units as belonging to a \emph{metro} area if their corresponding county population is greater than 1,000,000 ($n=396$). Similarly, we classify test units as belonging to a \emph{rural} area if the county population is less than 100,000 ($n=342$). It should be noted that these thresholds are not being applied at the county level. An accepted convention, when considering populations, is to refer to higher resolution ``tracks of land'' (e.g., municipalities)~\cite{censusMSA}. We choose our own thresholds that are consistent with general breakdown of the \gls{US} population: about 20\% of the \gls{US} population resides in rural areas. With our choice, we also have approximately 20\% of the dataset being classified as rural~\cite{metroRural}. Our choice of the metro threshold allows us to select the top 50 most populous metropolitan subset while also ensuring a similar sample size to the rural subset~\cite{censusMSA}.

It should be noted that there is no single generally accepted urban-rural classification. Our method differs from other (more established) designations such as~\cite{rural-classifications}. Exploring other classifications or using \glspl{MSA}~\cite{msa} is left for future work.

\subsection{Daily Patterns Preprocessing}

All measurement timestamps in the \gls{FCC} \gls{MBA} raw datasets use the \gls{UTC}. The unit profile dataset provides a timezone offset for each test unit. We used that information to convert \gls{UTC} timestamps to the test unit's local time. This step reduced the number of test units available for daily patterns analysis to approximately 3,000.

The reduction is caused by an incomplete unit profile dataset. Not all test units are included in the dataset, primarily because some of the information needs to be manually verified with the \glspl{ISP} and the unit profile dataset is published by the \gls{FCC} infrequently. Thus, we lack the timezone information for test units that are not included in the unit profile dataset and must exclude such test units from daily pattern analysis.

For some test units and months, a daylight saving time conversion should also have been performed. We omitted this step as it would not have changed our daily traffic patterns significantly.

For each month, we define the average downloaded ($\overline{D}$) and uploaded data ($\overline{U}$) as

\begin{equation}
    \overline{D} = \frac{1}{N} \sum_{i=1}^{N} (wired\_rx_i + wifi\_rx_i)
\end{equation}
\begin{equation}
    \overline{U} = \frac{1}{N} \sum_{i=1}^{N} (wired\_tx_i + wifi\_tx_i)
\end{equation}
where $N$ is the number of test units that kept sending measurements from January 2019 to June 2020, i.e., we focus on test units that have been running for over a year and can thus be considered stable. $wired\_rx$, $wifi\_rx$, $wired\_tx$, and $wifi\_tx$ refer to columns in the \textit{datausage} \gls{MBA} dataset.

\begin{figure*}[t]
    \centering
    \includegraphics[width=1.0\linewidth]{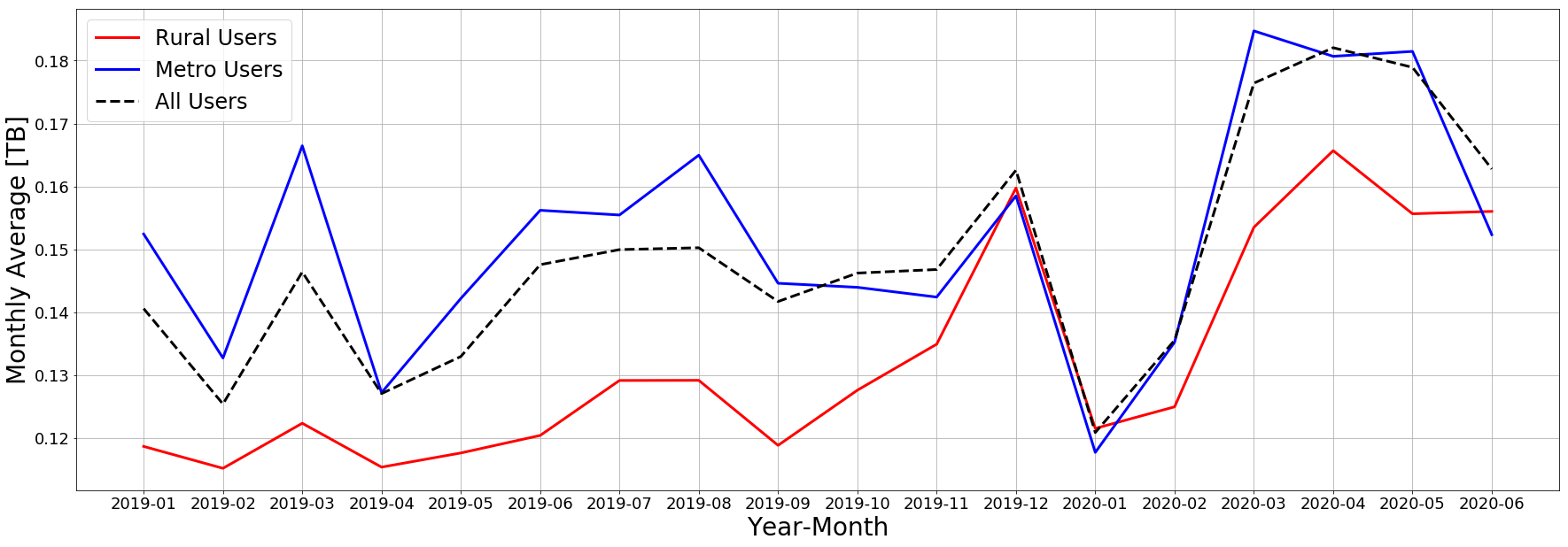}
    \caption{Graph showing the monthly average downloaded data per test unit from January 2019 to June 2020, partitioned by test unit county population.}
    \label{fig:downloadmetro_rural}
\end{figure*}

\begin{figure*}[t]
    \centering
    \includegraphics[width=1.0\linewidth]{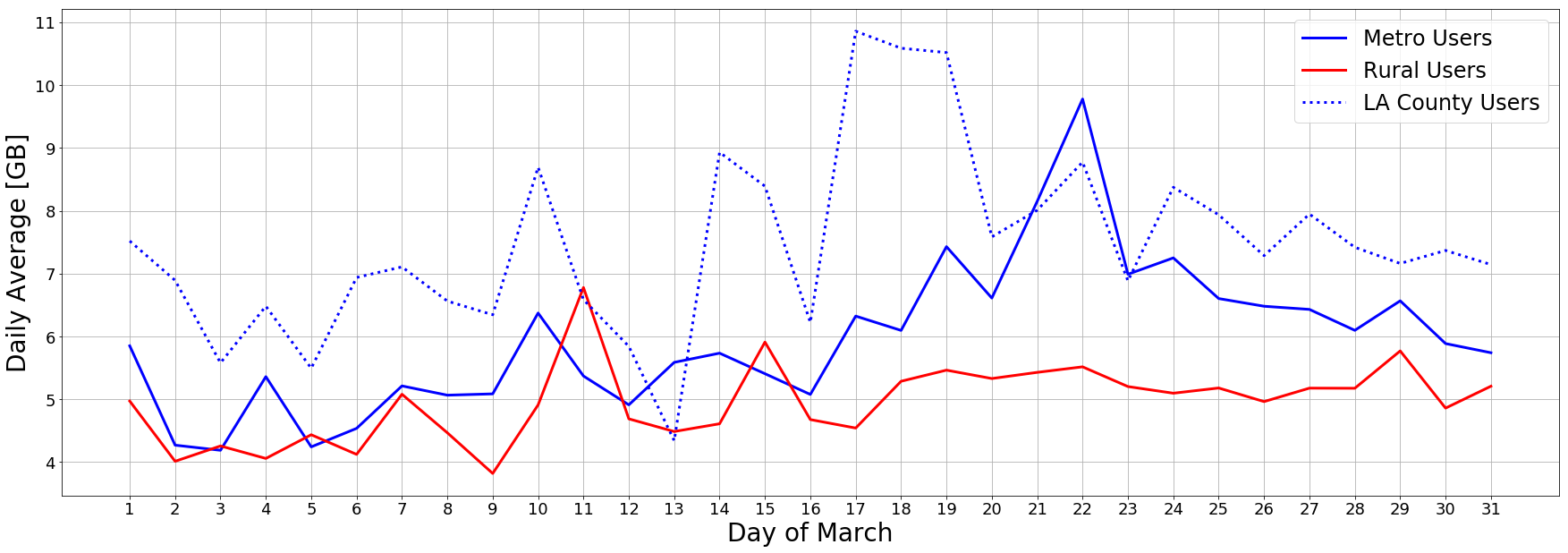}
    \caption{Graph showing the daily downloaded data per test unit in March 2020 for those in metro and rural areas, as well as Los Angeles (LA) county.}
    \label{fig:dailymetro_rural}
\end{figure*}

\section{Data Usage by Population Demographics}\label{sec:data-usage-by-population-demographics}

\subsection{Average Monthly Downloads}

A high-level overview of the effect of COVID-19 on fixed broadband internet usage can be seen by analyzing the average monthly downloaded data per test unit. As shown in \cref{fig:downloadmetro_rural}, we notice a sizable increase in overall downloaded data in the months of March, April, and May 2020. This increase is consistent across both metro and rural counties, although it is more pronounced in metro areas. The three pandemic months are the highest-usage months for metro areas. In contrast, rural areas only have one month larger than their pre-pandemic maximum of December 2019.

Interestingly, in the fourth month of the pandemic, we notice a drop in overall downloaded data in metro areas, while rural areas remain constant. This drop coincides with the relaxation of many stay-at-home orders, indoor dining bans, and curfews in cities across the \gls{US}~\cite{money2020la,gov2020nyc}.


\subsection{Average Daily Downloads}
Focusing on daily fixed broadband internet usage can allow us to understand the public's reactions to various events. In particular, by analyzing daily averages over the course of March, we can better understand the reactions to the various stay-at-home orders, school shutdowns, and curfews from an internet usage point-of-view.

To this end, we present in \cref{fig:dailymetro_rural} the daily average downloaded data usage for both metro and rural users. The first event we look to analyze is the White House first announcing its social distancing recommendations on March 16 for the entire country~\cite{trump2020coronavirus}. Following this announcement, we observe that users in metro areas use a significantly larger amount of data than those in rural areas. In the first half of the month, there was no such difference. It is unsurprising that those in cities and other metro areas reacted to this recommendation more drastically than those in less populated areas. Cities such as New York City, San Francisco, and \gls{LA} were among the hardest hit areas in the early weeks of the pandemic~\cite{cdc2020tracker}.

For many of the aforementioned cities, various levels of lockdowns were put in place (\cref{tab:state-lockdown}) on top of the White House recommendation. Therefore, we additionally investigate if we can correlate these local ordinances with county-level usage patterns. \gls{LA} county has the most test units in the \gls{FCC} \gls{MBA} dataset ($n=44$), so we examine the daily average downloaded data per test unit for all those residing in \gls{LA}.

In \cref{fig:dailymetro_rural}, we observe the first large spike on March 14, the day after many schools in the county shut down~\cite{haire2020LA}. March 14 was a Saturday, so this spike is not due to remote learning. However, if schools were shut down by this date, then we can reasonably expect that various other aspects of society were shut down. For example, although the official order to ban indoor dining came later, by March 14, many \gls{LA} restaurants were already closed~\cite{eater2020}. This would suggest that the spikes we observe following government orders are due to people seeking entertainment from the internet over outside options.

We also notice a larger spike in the days following the March 16 White House stay-at-home order.
\begin{figure*}[th]
    \centering
    \subfloat[\textbf{Pre-lockdown weekdays.} Average volume of downloaded data per test unit, broken down by the hour of the day, on weekdays in the pre-lockdown time period. \label{download-a}]{%
      \includegraphics[width=0.49\linewidth]{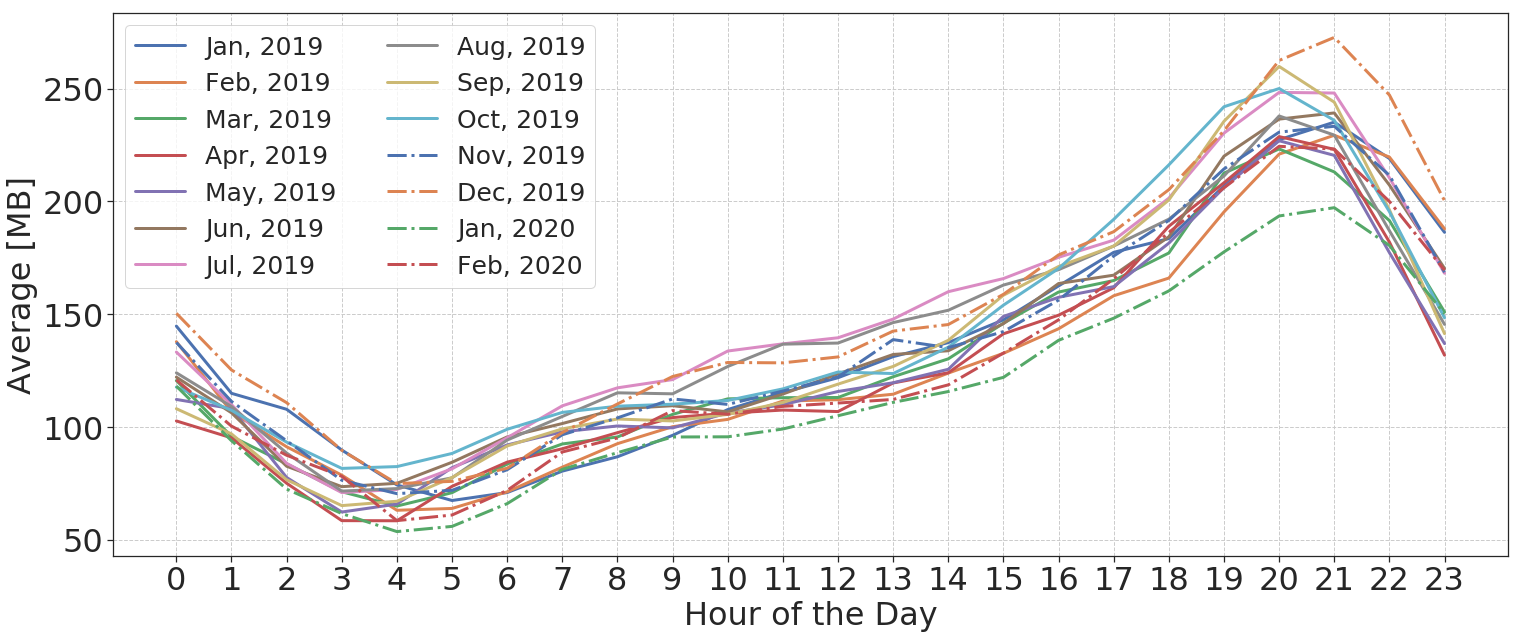}%
    }
    \hspace*{\fill}
    \subfloat[\textbf{Pre-lockdown weekends.} Average volume of downloaded data per test unit, broken down by the hour of the day, on weekends in the pre-lockdown time period. \label{download-b}]{%
        \includegraphics[width=0.49\linewidth]{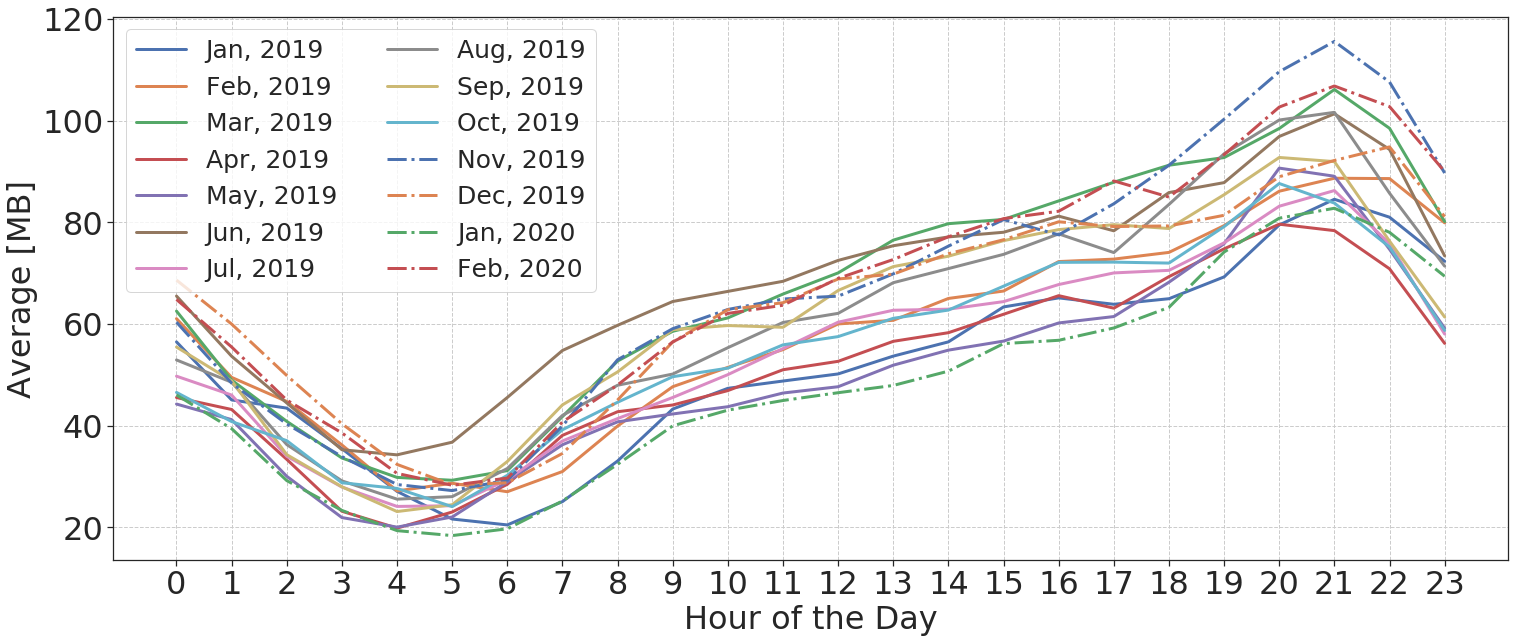}%
    }
    \\
    \subfloat[\textbf{Lockdown weekdays.} Average volume of downloaded data per test unit, broken down by the hour of the day, on weekdays in the lockdown time period. \label{download-c}]{%
        \includegraphics[width=0.49\linewidth]{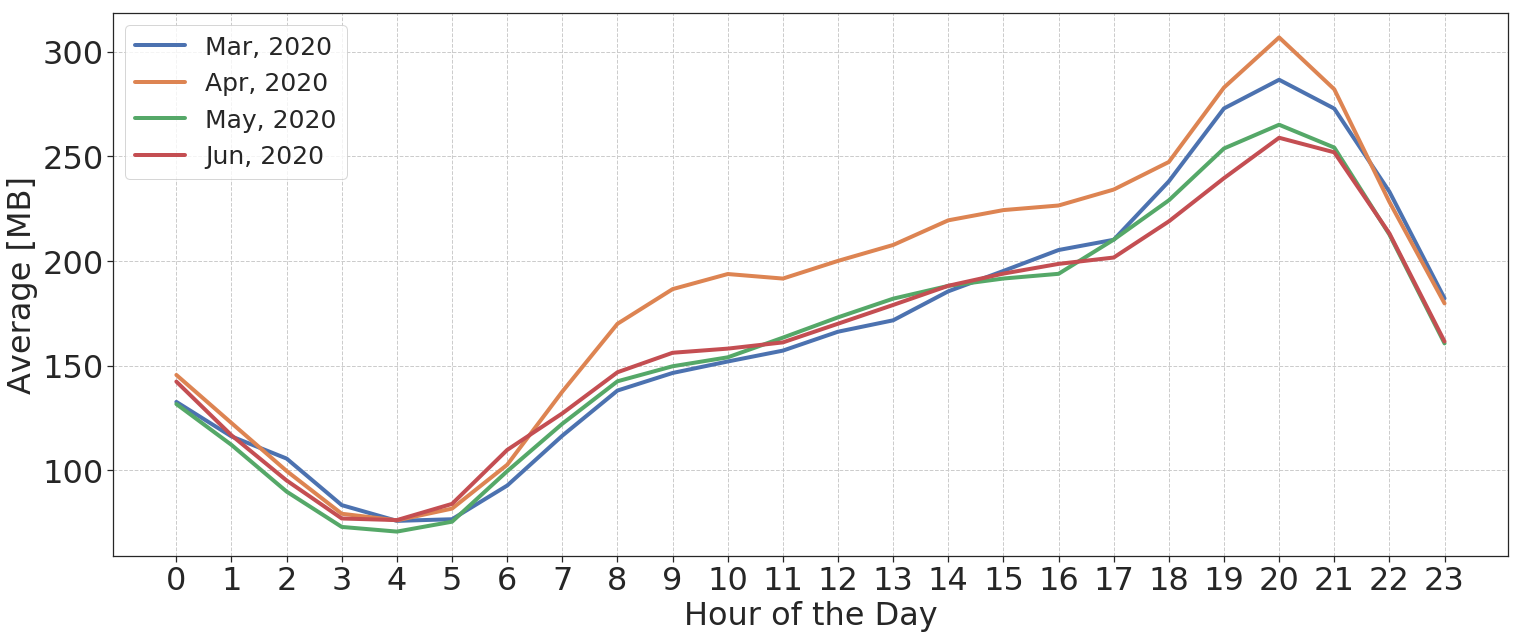}%
    }
    \hspace*{\fill}
    \subfloat[\textbf{Lockdown weekends.} Average volume of downloaded data per test unit, broken down by the hour of the day, on weekends in the lockdown time period. \label{download-d}]{%
        \includegraphics[width=0.49\linewidth]{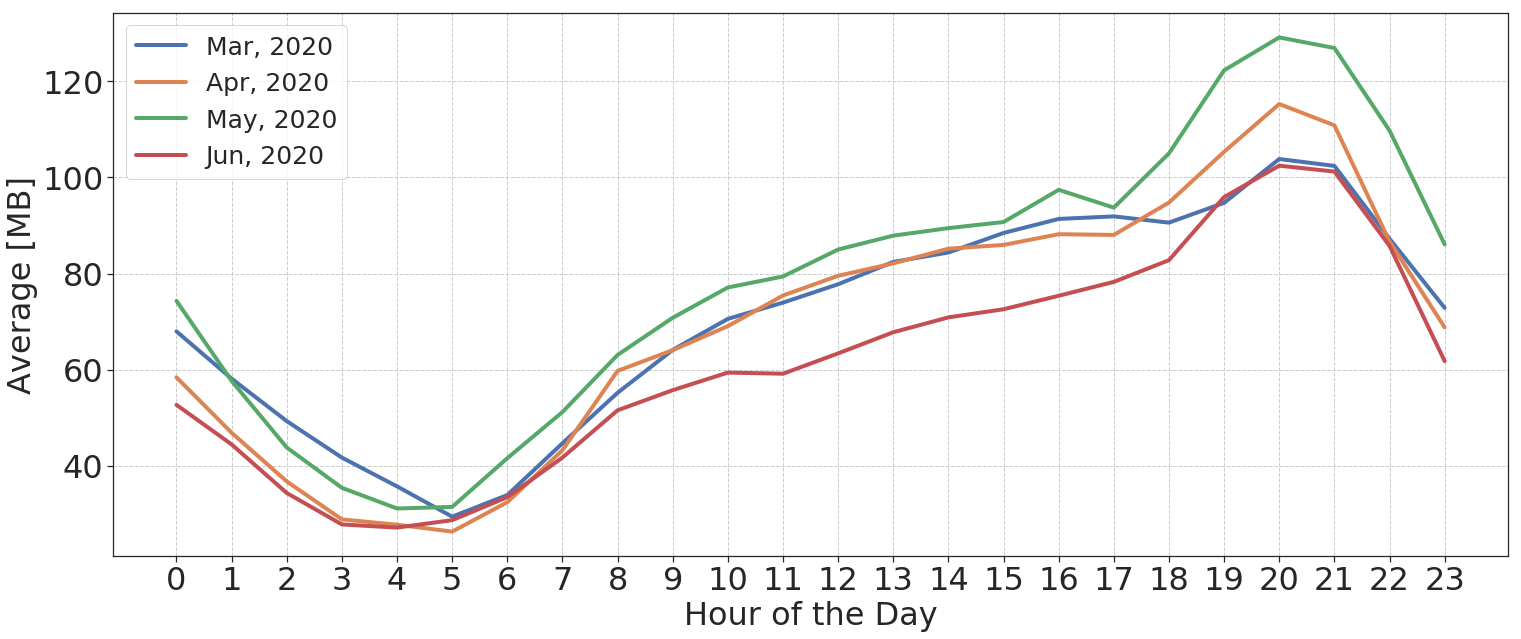}%
    }
    \\
    \subfloat[\textbf{2019 vs 2020 weekdays.} Average volume of downloaded data per test unit on weekdays in March-June compared between 2019 and 2020.
    \label{download-e}]{%
        \includegraphics[width=0.49\linewidth]{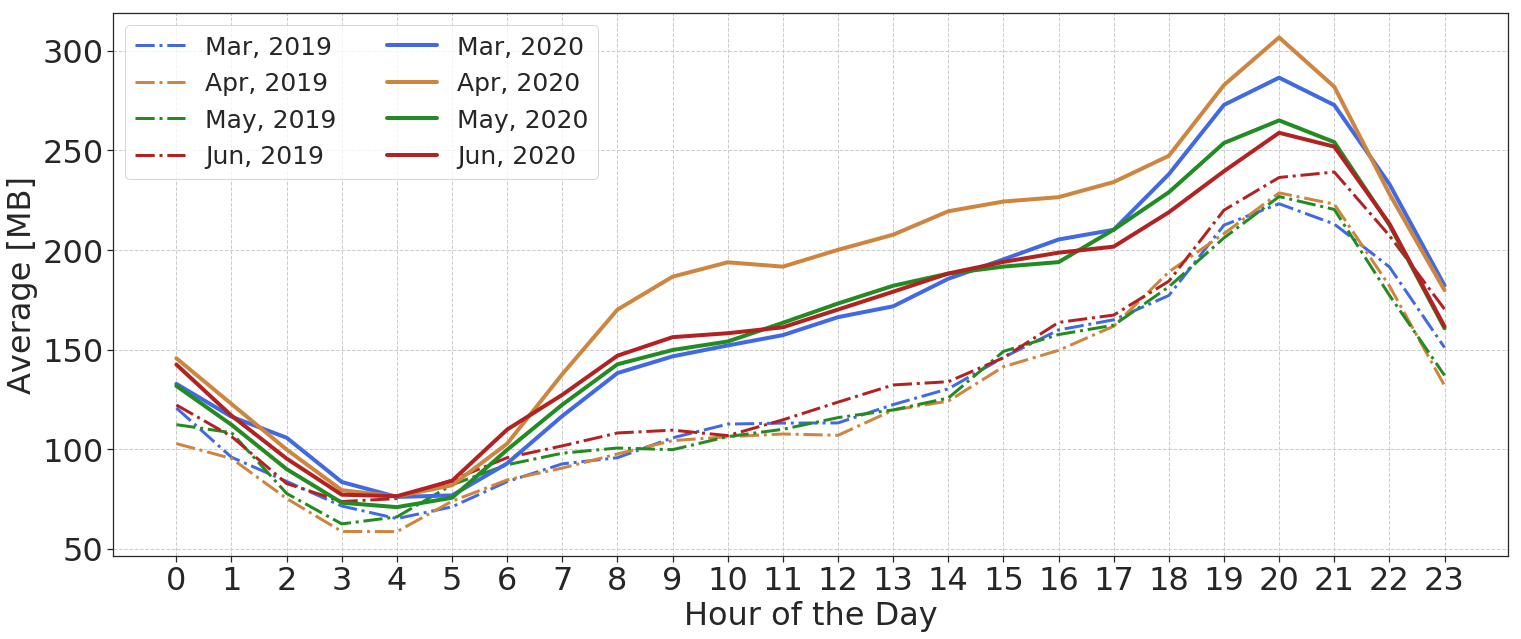}%
    }
    \hspace*{\fill}
    \subfloat[\textbf{2019 vs 2020 weekends.} Average volume of downloaded data per test unit on weekends in March-June compared between 2019 and 2020.
    \label{download-f}]{%
        \includegraphics[width=0.49\linewidth]{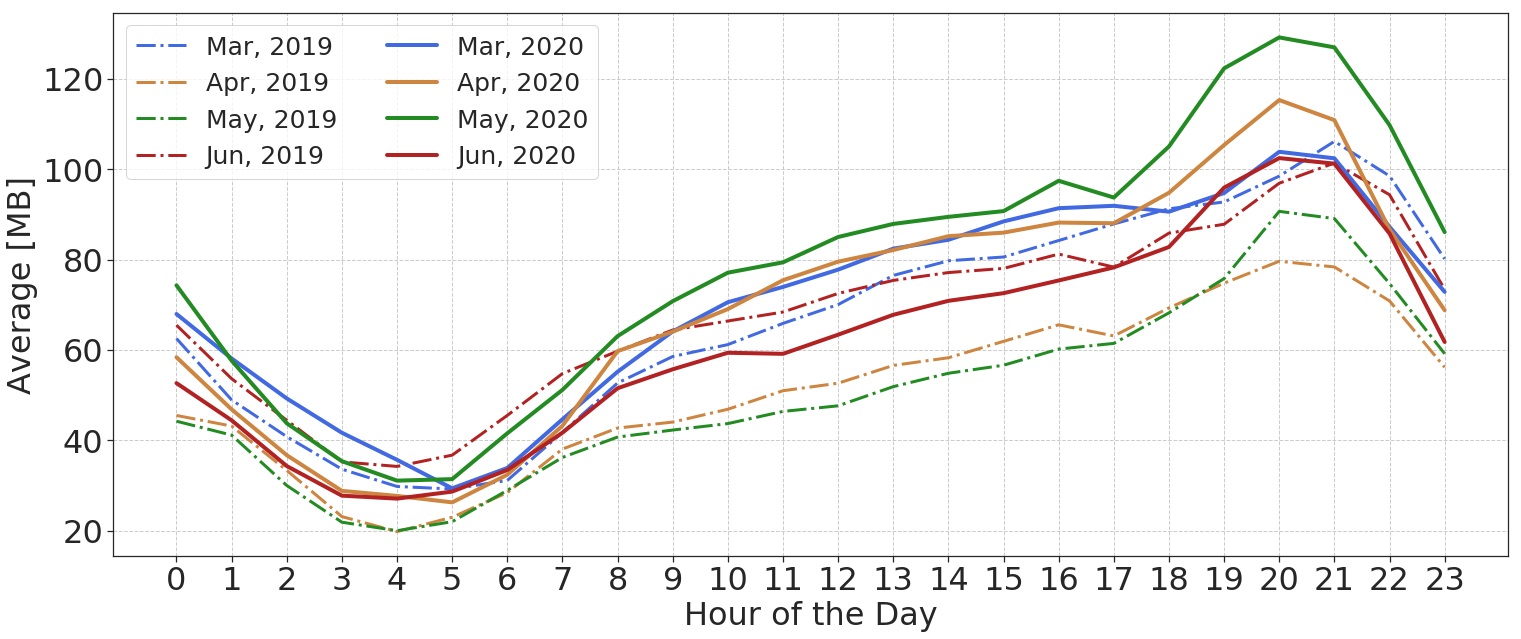}%
    }

    \caption{Graphs comparing and contrasting the average hourly downloaded volume of data per test unit in various time periods. The peak is always between 18:00 and 22:00, which is the internet rush hour. The volumes in the lockdown time period are generally larger than the volumes in the pre-lockdown time period. Lockdown weekday traffic patterns start to resemble weekend patterns.}
    \label{fig:download-data-per-user-hours-fig}
\end{figure*}

\begin{figure*}[th]
    \centering
    \subfloat[\textbf{Pre-lockdown weekdays.} Average volume of uploaded data per test unit, broken down by the hour of the day, on weekdays in the pre-lockdown time period.
    \label{upload-a}]{%
        \includegraphics[width=0.49\linewidth]{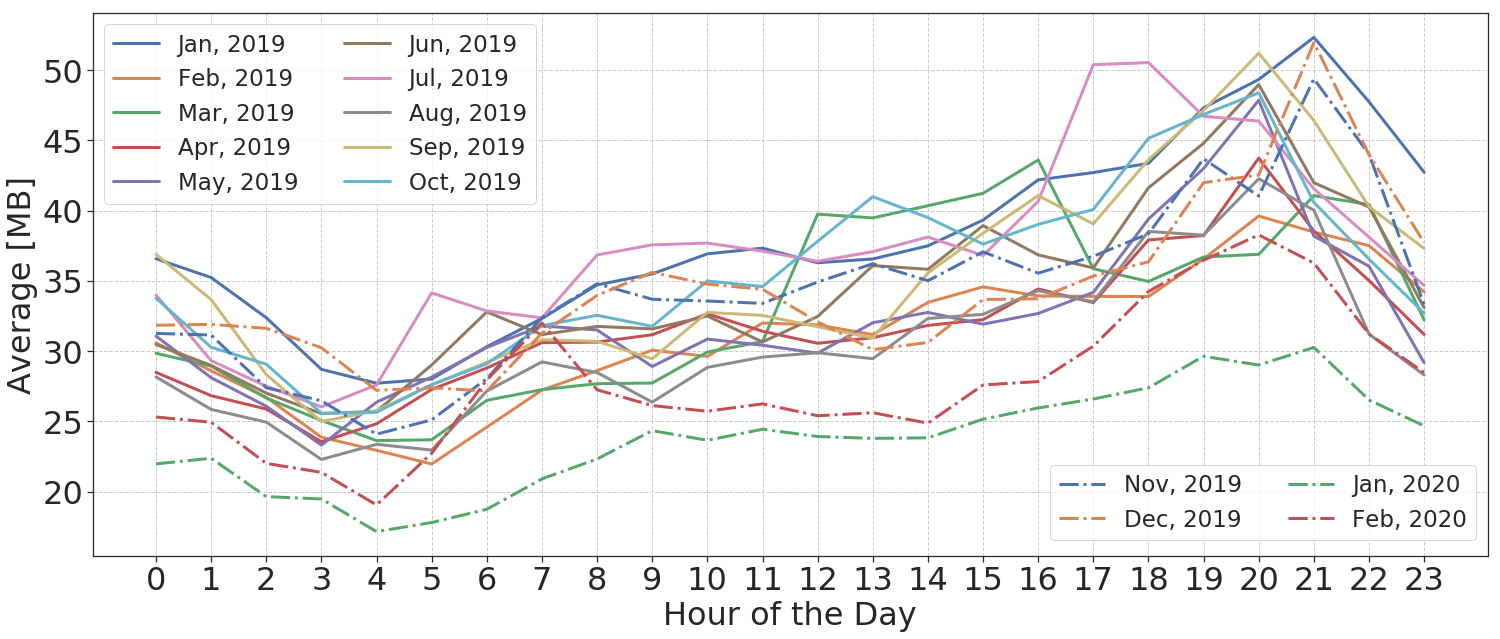}%
    }
    \hspace*{\fill}
    \subfloat[\textbf{Pre-lockdown weekends.} Average volume of uploaded data per test unit, broken down by the hour of the day, on weekends in the pre-lockdown time period.
    \label{upload-b}]{%
        \includegraphics[width=0.49\linewidth]{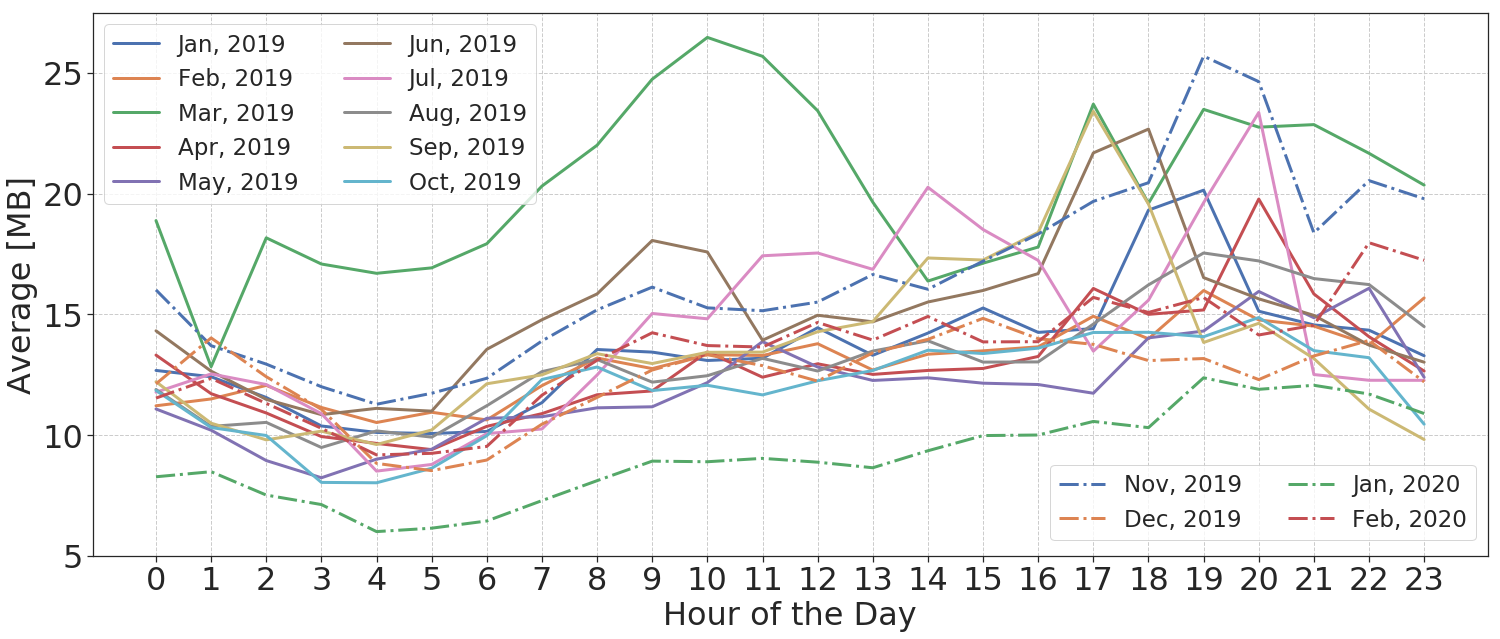}%
    }
    \\
    \subfloat[\textbf{Lockdown weekdays.} Average volume of uploaded data per test unit, broken down by the hour of the day, on weekdays in the lockdown time period.
    \label{upload-c}]{%
        \includegraphics[width=0.49\linewidth]{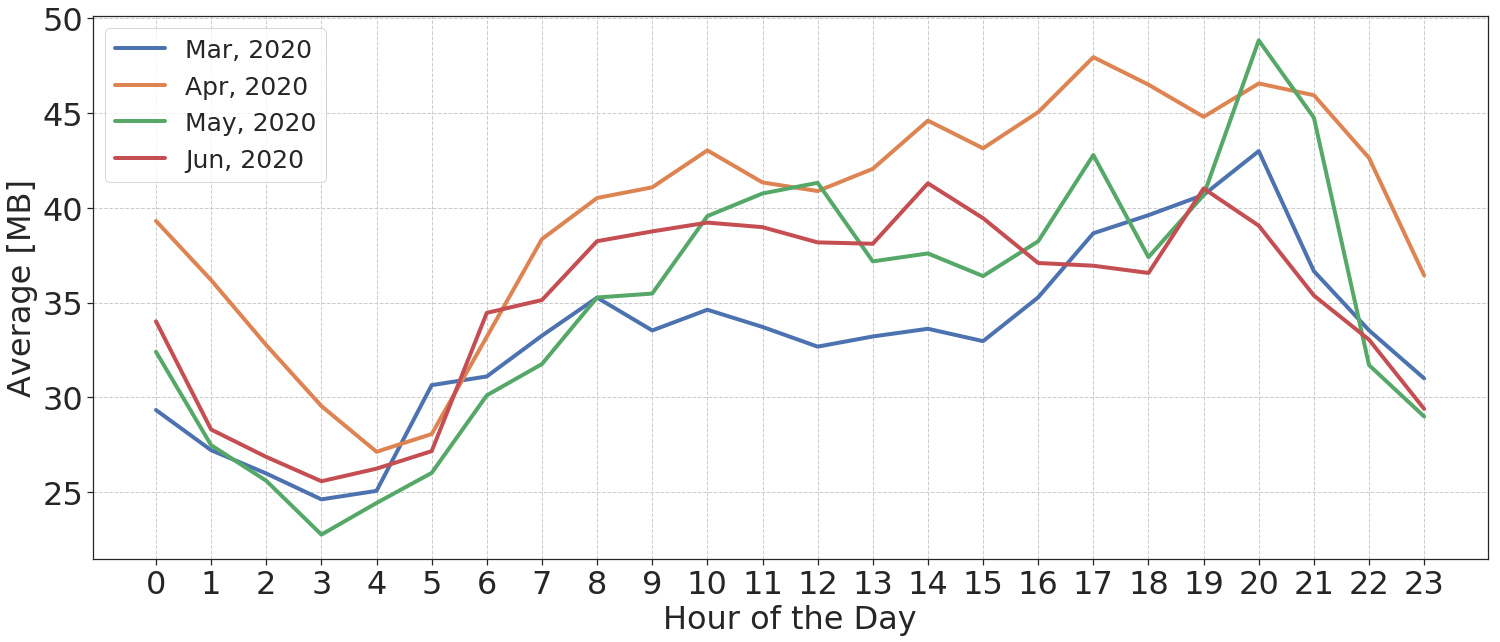}%
    }
    \hspace*{\fill}
    \subfloat[\textbf{Lockdown weekends.} Average volume of uploaded data per test unit, broken down by the hour of the day, on weekends in the lockdown time period. \label{upload-d}]{%
        \includegraphics[width=0.49\linewidth]{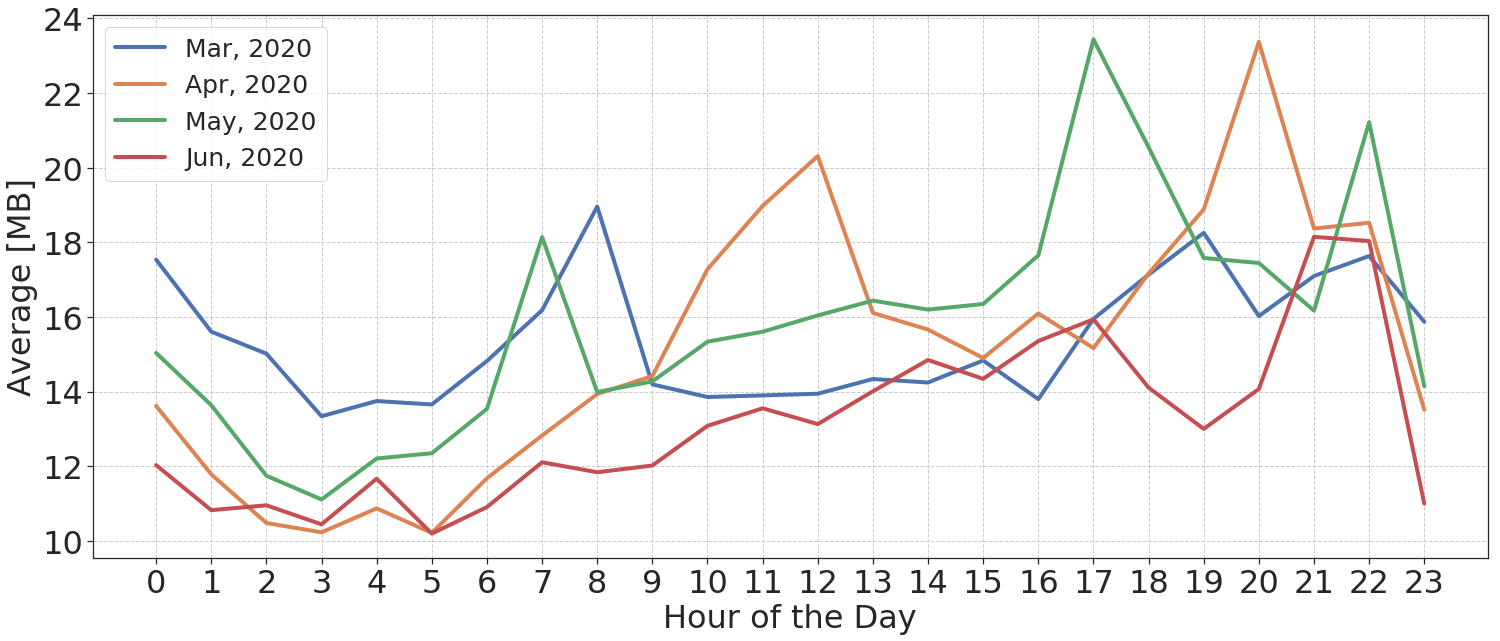}%
    }
    \\
    \subfloat[\textbf{2019 vs 2020 weekdays.} Average volume of uploaded data per test unit on weekdays in March-June compared between 2019 and 2020.
    \label{upload-e}]{%
        \includegraphics[width=0.49\linewidth]{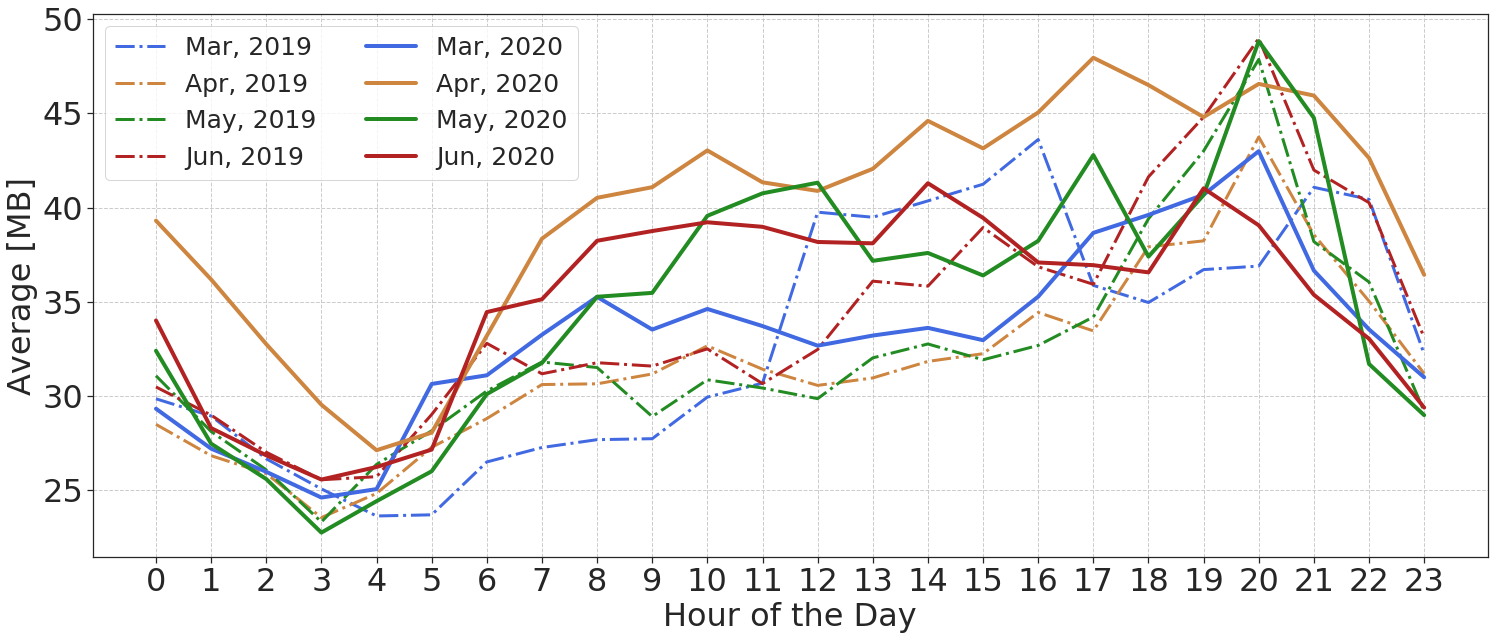}%
    }
    \hspace*{\fill}
    \subfloat[\textbf{2019 vs 2020 weekends.} Average volume of uploaded data per test unit on weekends in March-June compared between 2019 and 2020.
    \label{upload-f}]{%
        \includegraphics[width=0.49\linewidth]{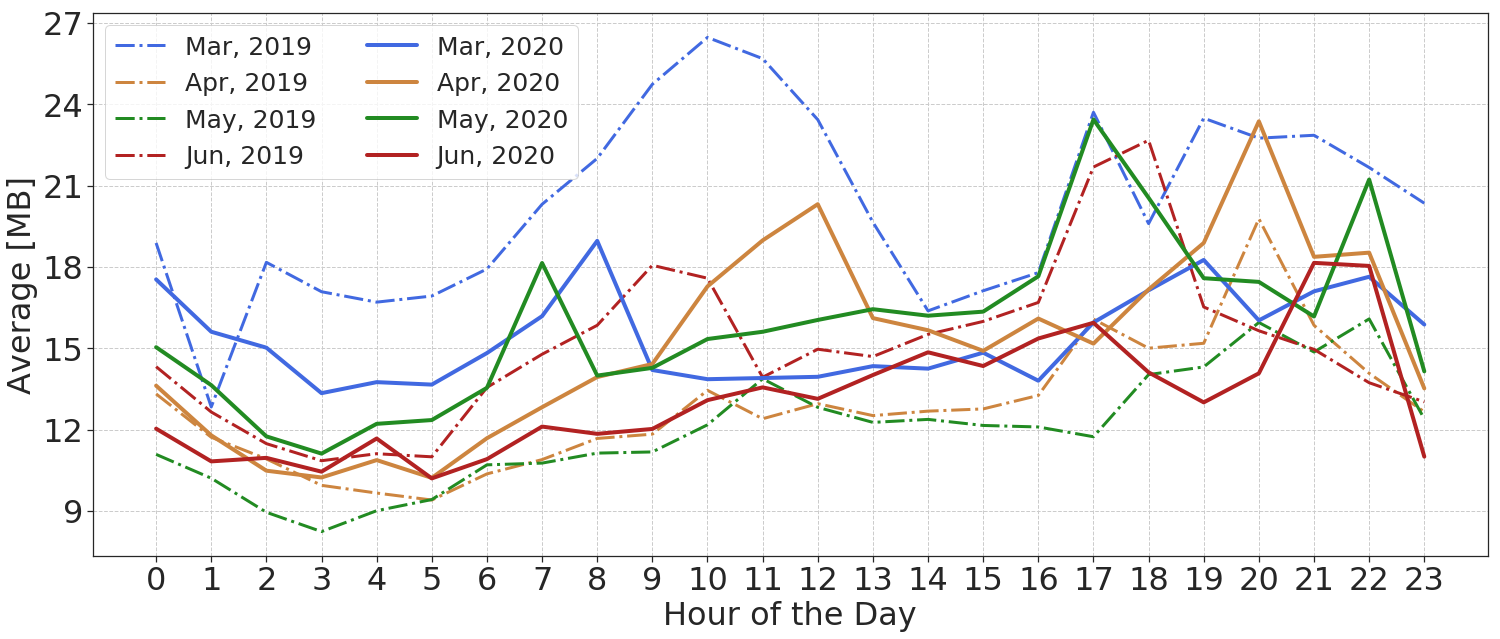}%
    }

    \caption{Graphs comparing and contrasting the average hourly upload volume of data per test unit in various time periods. Compared with download patterns, the lines in these graphs fluctuate more, especially on weekends. When comparing April and May between 2019 and 2020, we see an overall increase in uploaded data in 2020.}
  \label{fig:upload_data_per_user_hours_fig}
\end{figure*}

\section{Hourly Data Usage Patterns}\label{sec:hourly-data-usage-patterns}

Hourly data usage patterns (upload and download) could potentially provide insights into the behavior of fixed broadband internet users before and during the pandemic. Some applications such as music and video streaming generally increase the volume of downloaded data. Other applications such as \gls{VoIP}, video conferencing, or networked games influence the volume of uploaded data.

In this section, we study the average hourly data usage per test unit from multiple perspectives and attempt to correlate the observations with changing user behavior patterns during the lockdown. We contrast weekday and weekend patterns, analyze the differences in uploaded and downloaded traffic volumes, and compare with the data from the corresponding period of 2019 to eliminate potential seasonal differences.

The analysis in this section is primarily based on the \emph{datausage} table from the \gls{FCC} \gls{MBA} raw dataset. That table keeps track of the total number of bytes received and transmitted by all user's devices within each hour of the day. For hourly analysis, we needed to translate \gls{UTC} timestamps to the test unit's local time using the unit profile dataset, as described in \cref{sec:methodology}. All graphs in this section were generated with measurements from approximately 3,000 test units.

\subsection{Analysis of Download Patterns}\label{sec:analysis-of-download-patterns}

All graphs in \cref{fig:download-data-per-user-hours-fig} show a traffic peak between 18:00 and 22:00 hours, also known as the internet rush hour~\cite{internetrushhour}. Most likely explanation for this peak is post-workday internet multi-media consumption for entertainment purposes, e.g., Netflix or Youtube.

Graphs in \cref{download-a} and \cref{download-b} compare average weekday and weekend download traffic volumes in the pre-lockdown period, i.e., until end of February 2020. We clearly see that the overall average download volume is higher on weekdays. However, the distribution of activity throughout the day is different for weekdays and weekends. Between approximately 9:00 and 17:00, the weekend graph shows a relative increase in traffic volumes compared to the weekday graph and more variability month-to-month. This corresponds with the intuition that users typically use fixed broadband internet more during weekends when they are at home. The increase in variability can probably be partially explained by seasonal effects.

\cref{download-b} and \cref{download-d} compare weekend traffic before and during lockdown. We see that the minimum and maximum weekend daily traffic levels are roughly the same before and during lockdown. In March--May 2020 we see a significant increase in daytime weekend traffic volumes (green, yellow, blue in \cref{download-d}) relative to the period before lockdown. Interestingly, daytime volumes return to their normal pre-lockdown levels in June 2020. This can be partially explained by the fact that many restrictions in effect since April were being relaxed by June, or that users started spending more time outdoors, away from their home broadband internet connection. As a result, some users may have returned to spending their weekends outside of the home. 

When we compare lockdown weekdays with weekends in \cref{download-c} and \cref{download-d}, we see that the hourly weekday pattern starts to resemble weekend patterns. This trend is particularly visible for April 2020 (yellow). This trend is also clearly visible in \cref{download-e} which contrasts lockdown weekday data with weekday data from the same period of 2019. We clearly see that 2019 and 2020 curves cluster in two separate regions during daytime from roughly 8:00 to 18:00 hours. In other words, lockdown weekday download patterns morphed into weekend patterns. The same trend was observed by other authors~\cite{feldmann2020lockdown} in data from different vantage points.

It is worth nothing that in \cref{download-c} April 2020 appears to be an outlier. All other lockdown months appear to converge to similar traffic levels. Somewhat lower traffic levels in May and June 2020 can be probably partially explained by a new normal and relaxed lockdown restrictions. The traffic levels are highest in April because it is the first month in which most of the \gls{US} states were under a mandatory lockdown. Thus, this month represents the initial lockdown period during which a significant proportion of users was getting accustomed to working from home or online learning activities. The increase in evening traffic levels is likely due to social distancing restrictions \cite{lockdownsguide}, where people moved social activities to the internet.

On weekends (\cref{download-f}), we see a small traffic increase in March, and the increases in April and May are substantial. Weekend June 2020 daytime traffic levels saw a relative decrease compared with June 2019 levels, except for the evening spike. One way to explain the difference might be that after months of restrictions people try to spend weekends more outside their homes. This pattern coincides with the policies related to COVID-19 in the \gls{US}: the height of restriction was at the end of March and the beginning of April, and some restrictions were being relaxed in the mid-late May or June in many states~\cite{covid19restriction}.

We conclude that people use fixed broadband internet more and download more data because of COVID-19 and its related policies (e.g., lockdown, quarantine, work from home).

\subsection{Analysis of Upload Patterns}\label{sec:analysis-of-upload-patterns}

\cref{fig:upload_data_per_user_hours_fig} shows several plots of hourly average uploaded data. As expected, there is significantly more noise and fluctuation in these plots (particularly weekends), compared with the download plots in \cref{fig:download-data-per-user-hours-fig}.

We note that we have no explanation for the morning spike in March 2019 shown in \cref{upload-b}. Unlike the corresponding download patterns, hourly upload patterns before lockdown show little difference between weekdays and weekends, partly due to the noise.

 The average uploaded data levels April and May 2020 are larger than those in 2019, but in March and June, that is not the case (\cref{upload-e} and \cref{upload-f}). Possible reasons might be similar to those discussed earlier: the majority of \gls{US} states enforced peak lockdown restrictions at the end of March and the beginning of April. Some of the restrictions ended in mid-late May or June in some states~\cite{covid19restriction}.

 On weekdays, the daytime average uploaded data levels in April, May, and June 2020 are much larger than those in the corresponding months of 2019. On weekends, the increase is relatively smaller than on weekdays.

\begin{figure*}[t]
  \begin{minipage}[t]{0.490\linewidth}
    \centering
    \includegraphics[width=0.98\linewidth]{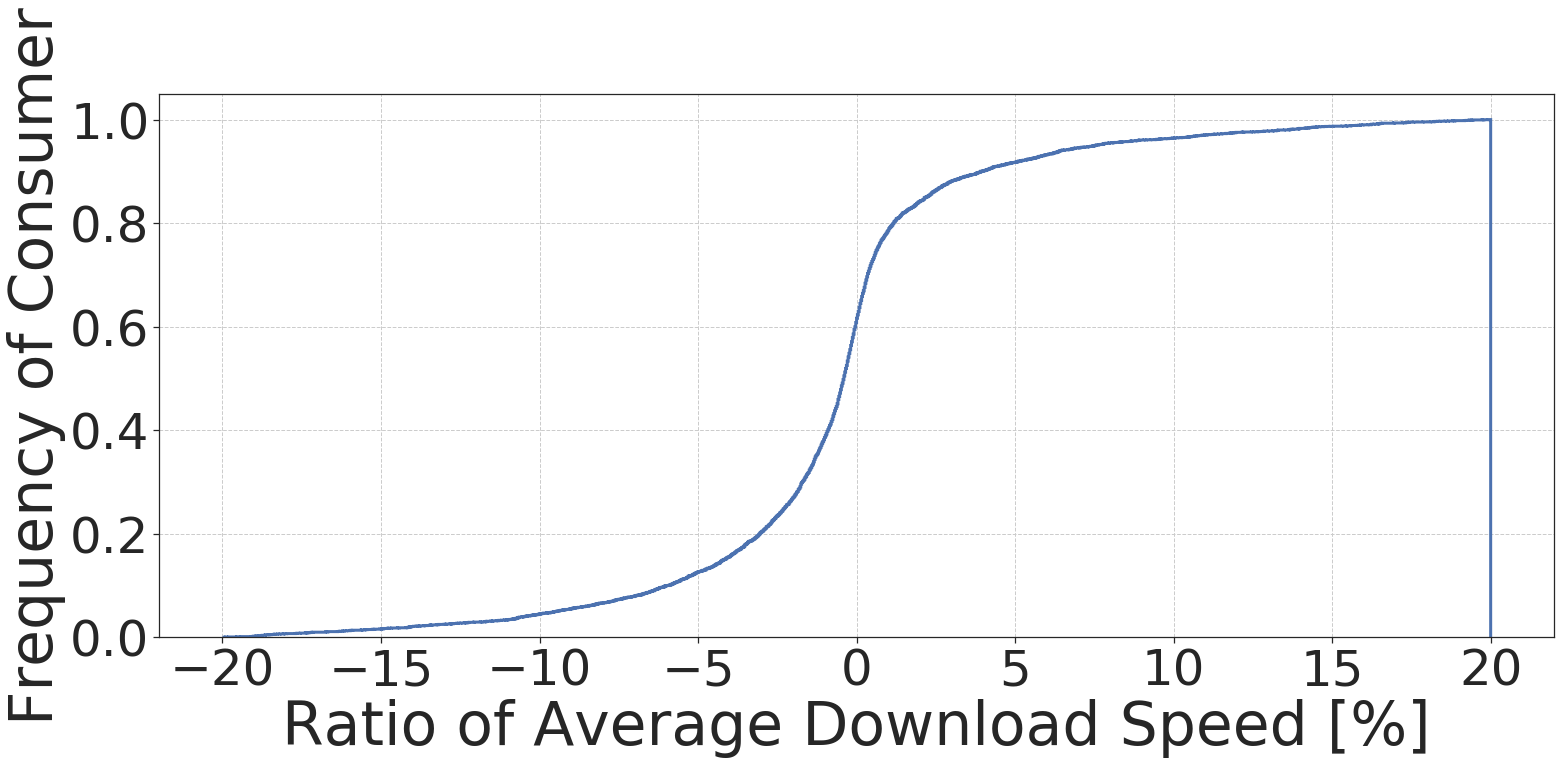}
    \caption{Cumulative distribution function (CDF) showing the ratio of average download speed per test unit between January and March of 2020.}
    \label{fig:downloadspeed2020}
  \end{minipage}
  \hspace*{\fill}
  \begin{minipage}[t]{0.490\linewidth}
    \centering
    \includegraphics[width=0.98\linewidth]{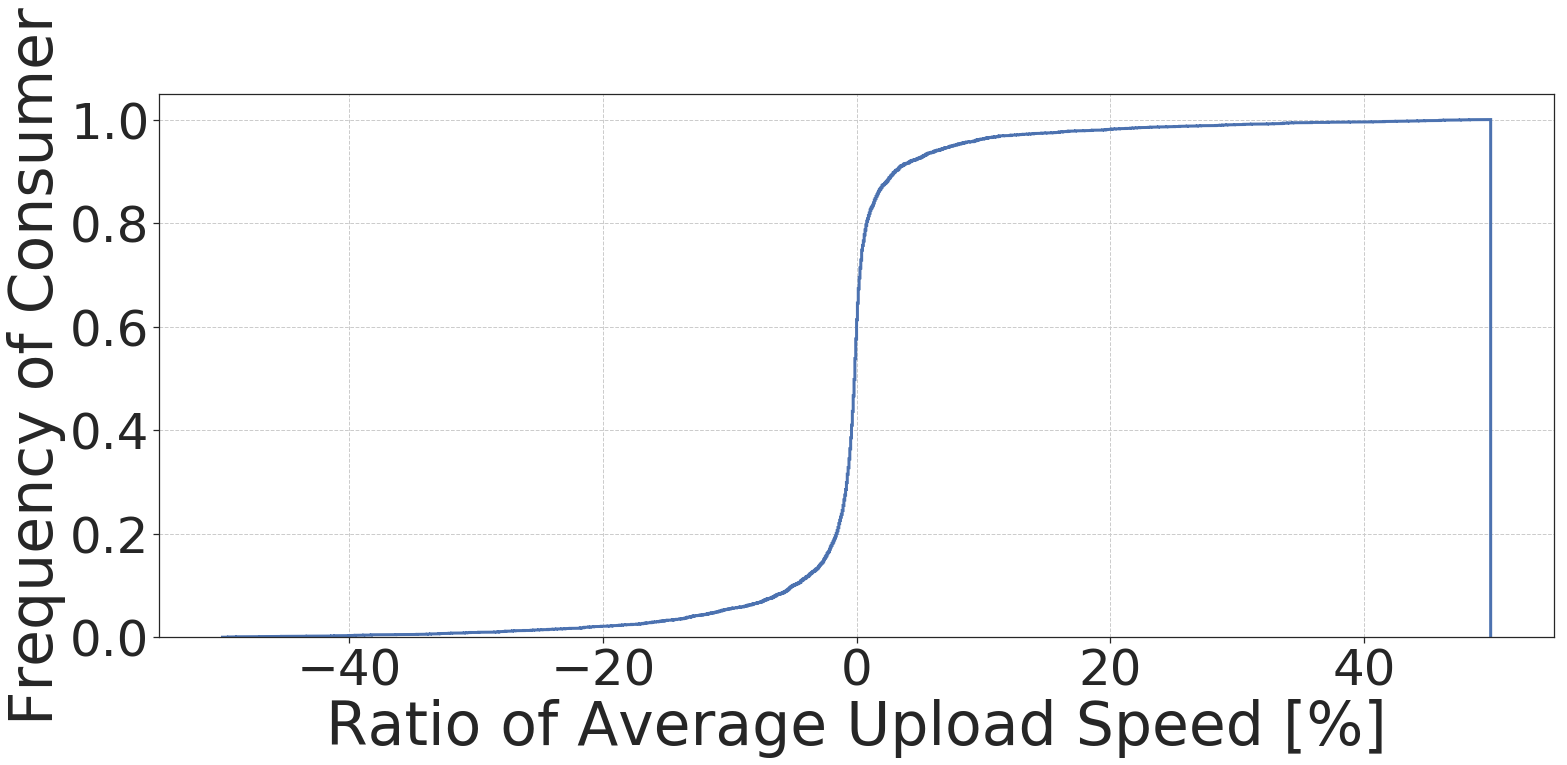}
    \caption{Cumulative distribution function (CDF) showing the ratio of average upload speed per test unit between January and March of 2020.}
    \label{fig:uploadspeed2020}
  \end{minipage}
\end{figure*}

\section{Average Speed}\label{sec:average-speed}

Given that internet data consumption generally increased across the \gls{US} in the patterns described in earlier sections~\cite{feldmann2020lockdown,liu2020characterizing}, an obvious question to ask if this increase led to changes in perceived network performance, e.g., due to increased network congestion. Knowing that the average volume of data per test unit increased significantly, one would expect the average internet upload or download speed to also change.

\subsection{Average Download Speed}\label{sec:average-download-speed}

In order to understand how the COVID-19 outbreak impacted the average download speed experienced by fixed broadband internet users, we calculate the ratio between the average download speed that users experienced between January and March 2020. Let $S_{\text{D}}$ be the download speed, we compute the average download speed ratio as follows:

\begin{equation}
\overline{S}_{\text{D}}\% = \frac{\overline{S}_{\text{D},\, \text{Mar}}-\overline{S}_{\text{D},\, \text{Jan}}}{\overline{S}_{\text{D},\, \text{Jan}}}\times 100
\end{equation}


\cref{fig:downloadspeed2020} shows the result in the form of a \gls{CDF}. We see that approximately 60\% of the test units experienced a decrease in their average download speed from January to March 2020. However, the decrease was not too significant, since for most of the test units, the decrease was smaller than 5\%. Therefore, while a significant number of test units experienced a decrease in average download speed as other studies suggest, not many experienced a severe decrease.

Cooper~\cite{cooper} argued that although some \glspl{ISP} were able to keep up with the increase in internet consumption, some cities felt the impact of the increase in data consumption leading users to experience a decrease in their download speed, e.g., the download speed in New York City decreased by 24\% in March.


\subsection{Average Upload Speed}\label{sec:average-upload-speed}

Similarly, we calculate the average upload speed ratio from January to March 2020. Let $S_{\text{U}}$ be the upload speed, we can calculate the average upload speed as follows:

\begin{equation}
\overline{S}_{\text{U}}\% = \frac{\overline{S}_{\text{U},\, \text{Mar}}-\overline{S}_{\text{U},\, \text{Jan}}}{\overline{S}_{\text{U},\, \text{Jan}}}\times 100
\end{equation}


\cref{fig:uploadspeed2020} shows the result. Approximately 70\% of test units saw their average upload speed decreasing from January to March 2020. As people transitioned most of their activities online, e.g., work from home, remote learning, and video conferences, an increase in uploaded data volumes was expected (as \cref{fig:uploadspeed2020} shows) and, consequently, a decrease in the average upload speed. Similarly to the situation observed with the download speed, while 70\% of test units experienced a decrease in their upload speed, this decrease was also modest, mostly less of 5\%.

One potential limitation of the analysis is that test units may get upgraded to higher \gls{ISP} service tiers in the middle of the analyzed time period. In that case, the performance of the broadband connection increases and, consequently, download and upload volumes may also increase.


\begin{figure*}[t]
  \begin{minipage}[t]{0.496\linewidth}
    \centering
    \includegraphics[width=0.98\linewidth]{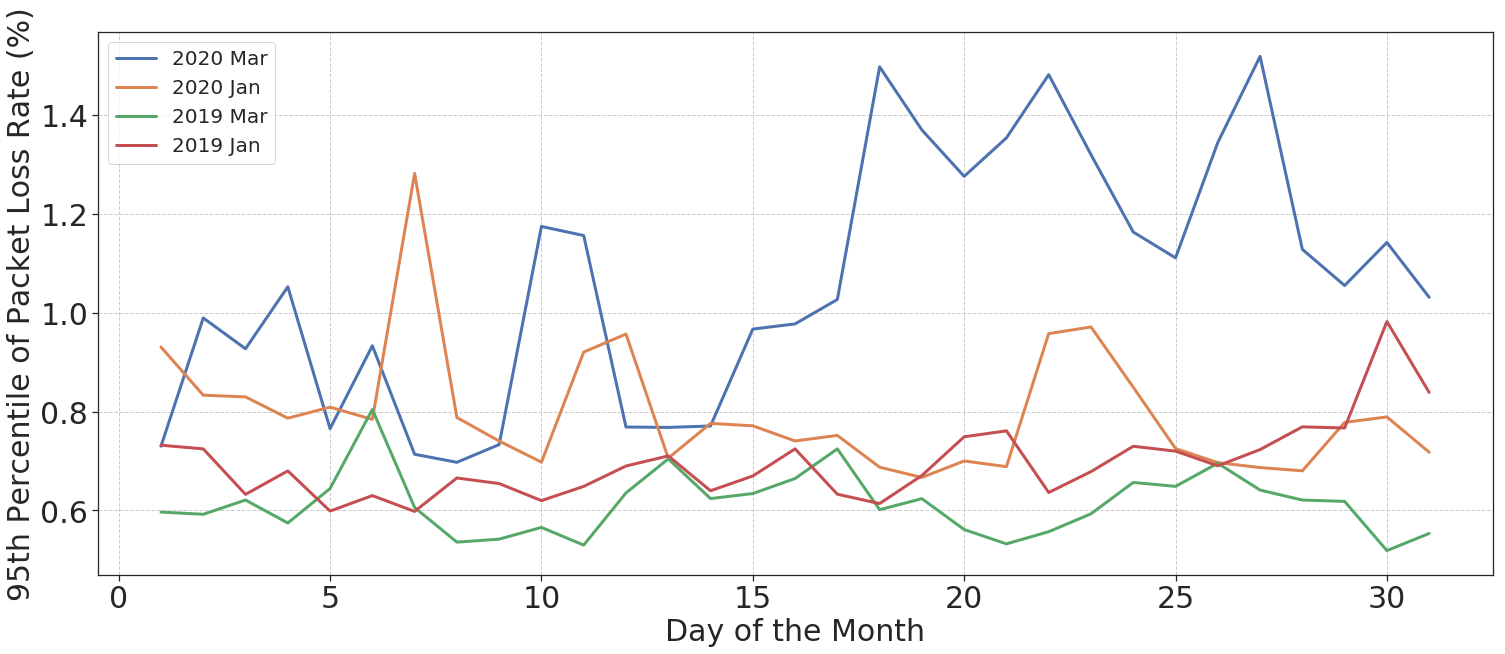}
    \caption{Graph comparing the 95th percentile packet loss per day for the months January, March 2019 and January, March 2020.}
    \label{fig:packetlossperday}
  \end{minipage}
  \begin{minipage}[t]{0.496\linewidth}
    \centering
    \includegraphics[width=0.98\linewidth]{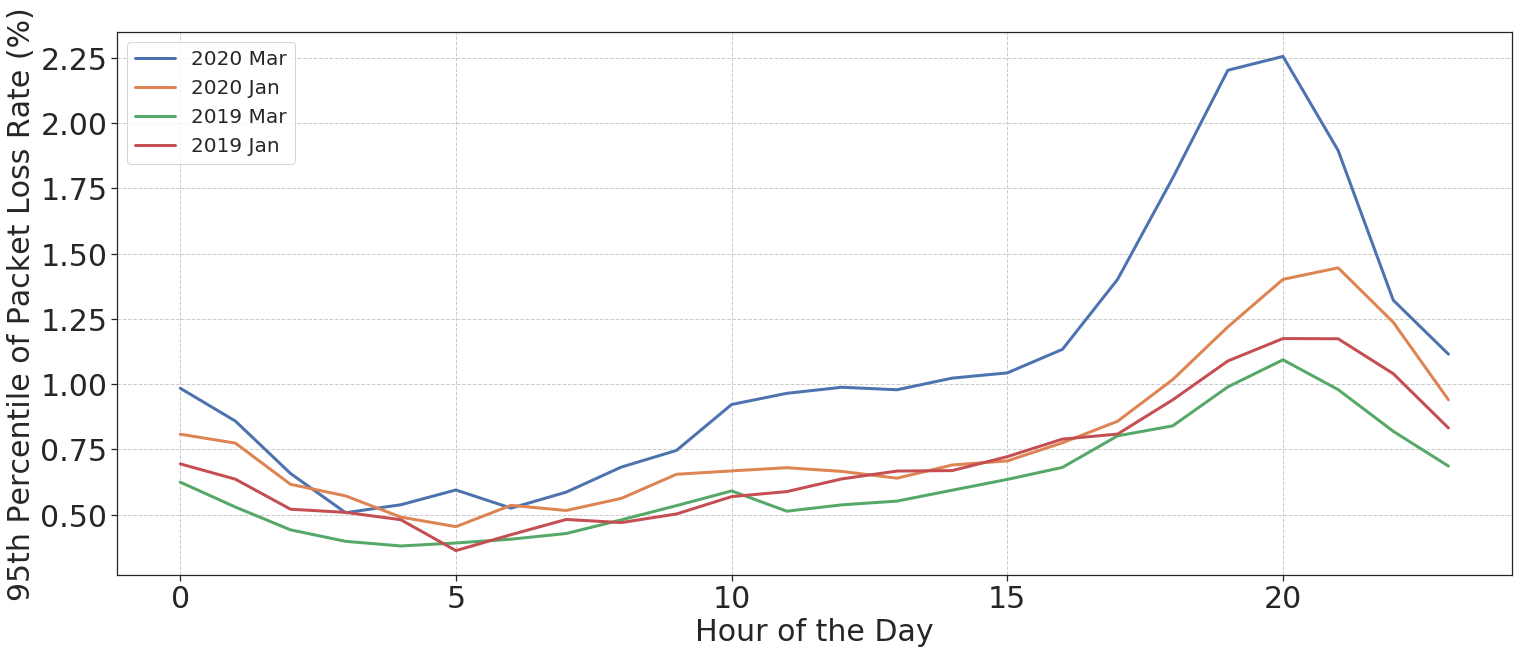}
    \caption{Graph comparing the 95th percentile packet loss per hour for the months January, March 2019 and January, March 2020.}
    \label{fig:packetlossperhour}
  \end{minipage}
\end{figure*}

\section{Packet Loss Analysis}\label{sec:packet-loss-analysis}

Apart from studying data usage and connection speed patterns, we can also examine the effects of COVID-19 restrictions on packet loss. With increasing fixed broadband internet usage during the lockdown period, network congestion is also likely to increase. Increasing network congestion will manifest itself in increasing packet loss.

Each test unit performs a \gls{UDP} latency measurement periodically throughout the day. During the measurement, the test unit sends a series of \gls{UDP} datagrams to a pre-defined set of servers. Parameters such as minimum, maximum, and average round  trip time are recorded, as well as the number of lost \gls{UDP} packets. The \gls{UDP} test is only performed when there is no user activity on the connection. Thus, the results are primarily influenced by the performance of the fixed broadband connection rather than local user traffic.

We use number the total number of sent and lost \gls{UDP} datagrams to calculate packet loss rate. To quantify the packet loss rate for one test unit, we define the metric as:

\begin{equation}
    L_\% = \frac{ F }{ F + S } \times 100
\end{equation}

where $L_\%$ is the packet loss rate, $F$ is the number of failed tests, and $S$ is the number of successful tests obtained from the \gls{MBA} \textit{udplatency} dataset.

Previous studies~\cite{kovacs} suggested that the majority of the \gls{US} did not experience any degradation of network performance. We thus focus on the subset of the test units that did experience network degradation and study packet loss at the 95th percentile, i.e., the 5\% of the users with the worst loss ratios.

\subsection{Daily Packet Loss}

\cref{fig:packetlossperday} shows the daily packet loss rate at the 95th percentile, i.e., the 5\% of the most ``lossy'' users,  over January and March 2019 and 2020. We see a significant increase in packet loss starting from mid-March 2020. Mid-March is the period when some \gls{US} states began imposing statewide lockdown orders. In this time period, the packet loss rate grows by 40--50\% compared to early March 2020, which is before the pandemic. Towards the end of March 2020, the packet loss rate was 80\% higher than that before the pandemic, implying that some network congestion occurred.

Since we do not observe similar trends in 2019, this suggests that this increase is not due to noise and might be instead due to the additional traffic generated by activities such as work from home, distance learning, and video conferencing. This increase also coincides with the increase in downloaded data we observed in metro areas in \cref{fig:downloadmetro_rural}.


\subsection{Hourly Packet Loss}

In this section, we examine the influence of the hour of the day on packet loss. \cref{fig:packetlossperhour} depicts the 95th percentile of packet loss over the course of a day in January, March 2019 and 2020. We found the packet loss ratio changing over the course of the day, with the maximum around 20:00. This is consistent with data upload and download patterns discussed earlier.

In March 2020, the packet loss rate increased steeply during daytime from 7:00 to 22:00, potentially due to the users' growing online activities like remote learning or work from home. However, this change was more evident during the evening. For example, we saw about 59\% growth from 18:00 to 22:00 when comparing March to January 2020. This time period lies outside of traditional working hours, suggesting that packet loss is most likely due to the increased traffic generated by recreational activities, rather than remote work or distance learning.

Our results are consistent with Candela et al.~\cite{Candela2020latency}, as they found the increase of packet loss might be affected by factors like the type of server, \gls{IP} version, and the time of the day.


\section{Discussion}\label{sec:discussion}

Because we were working with large datasets, \SI{300}{\giga\byte} of data in total from January 2019 to June 2020, one of the main challenges that we faced was to find an application that would be able to process all this data. We decided to use Google Cloud BigQuery and process all the data in Google Cloud \glspl{VM}. However, even in this configuration, some of the queries that analyzed the data over long time intervals took up to ten minutes to complete.

One of the biggest limitations in our analysis were outdated unit profile and unit census datasets. These datasets are published by the \gls{FCC} as part of the \gls{MBA} raw data. While the raw \gls{MBA} datasets are published regularly, unit profile and unit census datasets are only updated when a new annual report is published. This is understandable since some of the information in those datasets need to be manually cross-validated. At the time of writing, the 9th \gls{FCC} \gls{MBA} report was the most recent report. That report covered September-October 2018, hence our most recent unit profile and unit census datasets were from 2018. Having more recent unit profile and unit census datasets would have allowed us to perform more accurate demographic and geography analysis.

For the purpose of assessing users' behavior from network traffic patterns, we were limited by the fact that the \gls{FCC} \gls{MBA} datasets do not provide application or service level information about the traffic. Thus, we were not able to examine the relative contribution of various types of applications (\gls{VoIP}, YouTube, video conferencing) to traffic patterns. Others have examined such data in the context of COVID-19 related lockdowns, e.g., the authors in~\cite{bottger2020internet}.

When used in isolation, the \gls{FCC} \gls{MBA} dataset provides only a limited insight into how households deal with short-term and long-term crises, e.g., natural disasters or epidemics. The reasons include: 1) the lack of application-layer detail, i.e., what network applications are being used and when; 2) sampling bias; and 3) inability to correlate fixed broadband measurement data with similar measurements from other related vantage points, e.g., \glspl{ISP}, network operators, or mobile networks.

For the purpose of emergency response and planning, it might be beneficial to include a mechanism in the \gls{FCC} \gls{MBA} dataset to track application usage in a privacy-respecting manner. For example, the test unit could classify uploaded and downloaded traffic into a few broad classes such as video stream, real-time communication, \gls{VPN}, and others. On a larger scale, such information might provide useful insight into the situation in a particular geographic area, for example, during a natural disaster.

Given that the \gls{FCC} \gls{MBA} program relies on volunteers, the collected data is likely to exhibit sampling bias. The dataset does not attempt to balance by rurality or subscriber counts. That is, the test units are more likely to be run by users somewhat technically savvy, more likely to reside in metro areas, or using a specific kind of broadband technology. This bias prevents us from extrapolating trends to the entire population of \gls{US} fixed broadband users. Thus, we can only discuss trends with respect to the population of test units.


Finally, other network measurements exist that might provide complementary information to the \gls{FCC} \gls{MBA} dataset. For example, during the pandemic, Google published regular mobility reports~\cite{google-mobility-report} based on data collected from Android devices. The authors in~\cite{lutu2020characterization} show the level of mobility detail that could be obtained from mobile network operator in a privacy-respecting manner. For disaster planning and recovery purposes, it might be beneficial to find a way to combine these and the \gls{FCC} \gls{MBA} dataset.

\section{Conclusion}\label{sec:conclusion}

We studied the data collected through the \gls{FCC} \gls{MBA} program from January 2019 to June 2020. Our goal was to understand the performance of fixed broadband internet in the \gls{US} before and during the COVID-19 pandemic, and also to look for clues about changes to user behavior based on their data usage patterns. We attempted to look at the data from multiple perspectives: by geography and population size, by weekdays and weekends, and over the course of the day.

Overall, we found a significant increase in fixed broadband data usage consistent with the lockdown timeline. Our observations are more or less consistent with the observations that others made using data from different vantage points (e.g., \glspl{ISP}, \glspl{IXP}, network service providers). We reviewed and summarized some of the existing work in \cref{sec:related-work} for context.

Broken down by geography and population density, our analysis shows traffic spikes in both rural and metro areas. However, the rural traffic spike is initially shorter and less pronounced. Towards June 2020, we see comparable traffic level values in metro and rural areas. Our data suggest that users in metro areas react to government recommendations more rapidly than users in less populated areas.

We further attempted to correlate local ordinances with county-level usage patterns for Los Angeles county. There, we saw the first traffic spike on March 14, a day after many schools in the county shut down. Another spike came after the March 16 White House stay-at-home order. We note that our geography and demographic analysis is partially limited by older (2018) unit profile and unit census datasets and fairly small number of test units at the county level.

Looking at weekday and weekend traffic patterns, we saw a significant increase in daytime weekend traffic from March to May 2020. By June the levels return to their normal pre-lockdown levels, suggesting that users might have complied with stay-at-home recommendations only initially. Consistent with others, we see the weekday daytime traffic pattern changing into the weekend pattern, with a significant increase of uploaded and downloaded data during the day. We attribute most of that increase to distance learning and work from home.

Our analysis of the average hourly downloaded and uploaded data indicates that the policies related to COVID-19, such as lockdown, stay-at-home order, work from home recommendation, increase the usage of internet in the \gls{US} on both weekdays and weekends.

Our average speed analysis uncovered a slight 5\% decrease in average speed for about 60-70\% of test units. We also found a significant, almost 50\%, increase in packet loss for some test units starting mid-March 2020.

\subsection{Future Work}\label{sec:future-work}

The results shown in this paper present various potential future avenues for exploration.

In our population demographics analysis, we had used county-level populations as a measure of population density and custom population thresholds to classify test units into urban and rural. Our method is designed to classify approximately 20\% of the test units as rural~\cite{metroRural}.

The \gls{FCC} unit census dataset provides a census block group for most test units. The area and population of each census block group can be obtained from publicly-available sources. Thus, the population density for most test units can be either calculated, or obtained in the form of a data layer for ArcGIS (a popular geographic information system)~\cite{arcgis} from the \gls{EPA} website.

It should probably be noted that there is no single generally accepted urban-rural divide. Our classification differs from other (perhaps somewhat more established) designations such as~\cite{rural-classifications}. Exploring other classifications or using \glspl{MSA}~\cite{msa} is left for future work.

Having a more recent unit profile and unit census data could increase the test unit sample size used in the analysis. Since most of our analysis does not rely on the cross-validated \gls{ISP} information included in the unit profile dataset, obtaining a more recent partial (without cross-validated \gls{ISP} information) unit profile data from the \gls{FCC} would help improve the analysis.

For user behavior analysis, the biggest limitation of our approach is that we do not know how specific applications contribute to the overall traffic for each test unit, e.g., Netflix, YouTube, Zoom. Incorporating other datasets with this information obtained from other vantage points could help us better understand user behavior from network traffic patterns.

Interesting insights into user behavior during lockdown might be perhaps obtained by correlating fixed and mobile broadband internet usage. These two datasets, when used together, would present a more complete picture of user behavior.

We have only analyzed a subset of the metrics provided by the \gls{FCC} \gls{MBA} dataset. Our study could be expanded by analyzing metrics such as latency, \gls{VoIP} and media streaming performance, \gls{DNS} performance, and others. Additionally, studying network performance by broadband connection type (cable, fiber, satellite) might provide additional insights.

\section*{Acknowledgment}\label{sec:acknowledgment}
\addcontentsline{toc}{section}{Acknowledgment}

Jessica De Oliveira Moreira was funded by Craig Newmark Philanthropies while working on the project. She would also like to thank the \gls{DIMACS} REU Summer 2020 Research Program and the Barnard Computer Science Department for additional support.

\bibliographystyle{IEEEtran}
\bibliography{bibs/references}
\end{document}